\documentclass{revtex4}

\usepackage{latexsym}
\usepackage{graphics}
\usepackage{amsmath}
\usepackage{amsfonts}
\usepackage{dsfont}
\usepackage{amssymb}

\newtheorem{teo}{Theorem}[section]
\newtheorem{lem}[teo]{Lemma}
\newtheorem{corol}[teo]{Corollary}

\newtheorem{prop}[teo]{Proposition}
\newtheorem{rem}[teo]{Remark}

\newcommand\e{{\rm e}}
\newcommand\p{{\partial}}
\newcommand{\beq}{\begin{equation}}
\newcommand{\eeq}{\end{equation}}

\newcommand{\N}{{\mathds{N}}}
\newcommand{\Z}{{\mathds{Z}}}
\newcommand{\R}{{\mathds{R}}}

\newcommand{\C}{{\mathds{C}}}

\newcommand{\HH}{{\mathds{H}}}

\renewcommand{\H}{{\mathcal{H}}}
\newcommand{\D}{{\rm dom}}
\newcommand{\Ran}{{\rm ran}}
\newcommand{\A}{{\mathcal{A}}}
\renewcommand{\a}{{\mathsf{a}}}
\renewcommand{\u}{{\mathsf{u}}}
\renewcommand{\v}{{\mathsf{v}}}
\newcommand{\ee}{{\mathsf{e}}}
\newcommand{\del}{{\mathsf{d}}}

\newcommand{\sgn}{{\rm sgn}}
\renewcommand\det{{\rm det}}
\renewcommand{\Re}{{\rm Re}}
\renewcommand{\Im}{{\rm Im}}
\newcommand{\Sp}{{\rm Sp}}
\newcommand{\Tr}{{\rm Tr}}
\newcommand{\F}{{\mathcal{F}}}

\DeclareMathOperator*{\Rz}{Res_0}
\DeclareMathOperator*{\Ru}{Res_1}

\DeclareMathOperator*{\Rk}{Res_k}
\DeclareMathOperator*{\ctg}{ctg}

\begin{document}
\title{Singular perturbations with boundary conditions\\ and the Casimir effect in the half space}

\author{S. Albeverio$^{(1)}$, G. Cognola$^{(2)}$, M. Spreafico$^{(3)}$, S. Zerbini$^{(2)}$}

\affiliation{(1) Inst. Appl. Math., HCM, IZKS, SFB611 University of Bonn,  
Wegelerstr. 6, 53115 Bonn, Germany\\ \small
and BiBoS (Bielefeld-Bonn), CERFIM (Locarno); 
Accademia di Architettura (USI, Mendrisio);\\ 
Dipartimento di Matematica, Universit\`a di Trento.\\
(2) Dipartimento di Fisica, Universit\`a di Trento,  Povo 38100, Italy,
and INFN, Gruppo Collegato di Trento.\\ 
(3) Dipartimento di Matematica, Universit\`a di Trento,  Povo 38100, Italy.\\
On leave from ICMC, Univer\-sidade de S\~{a}o Paulo, S\~{a}o Carlos, Brasil.
}

\begin{abstract}
We study  the self adjoint extensions of a class of non maximal multiplication 
operators with  boundary conditions. 
We show that these extensions correspond to singular rank one perturbations 
(in the sense of \cite{AK}) of the Laplace operator, 
namely the formal Laplacian with a singular delta potential, on the half space.
This construction is the appropriate setting to describe the Casimir effect 
related to a massless scalar field in the flat space time 
with an infinite conducting plate and in the presence of a point like "impurity". 
We use the relative zeta determinant (as defined in \cite{Mul} and \cite{SZ}) 
in order to regularize the partition function of this model. 
We study the analytic extension of the associated relative zeta function, 
and we present explicit results for the partition function, 
and for the Casimir force.
\end{abstract}

\maketitle


\def\ot{o\left(N^{\left[\frac n2\right]-1}\right)}
\def\la{\lambda}
\def\al{\alpha}
\def\gna{\frac{a_n}\al+b_n}
\def\rl{ \sqrt\lambda \sgn(\Im\sqrt\lambda) }

\def\cUno{\frac{\pi}{\sqrt2 \al}+\frac{\pi}{\sqrt2}-\frac{i\pi}{\sqrt\la}}
\def\cDue{\frac{\pi^2}{2\al}+\pi\log(-\la)}
\def\cTre{\sqrt2\pi^2\left(\frac1\alpha-1-i\sqrt{2\la}\right)}

\def\trUno{\frac{i}{2\la^{3/2}} \frac{1}{\frac{1}{\sqrt2 \al}+\frac{1}{\sqrt2}-\frac{i}{\sqrt\la}}}
\def\trDue{\frac{\pi}{\la} \frac{1}{\cDue}}
\def\trTre{\frac{1}{i\sqrt{2\la}\left(\frac1\alpha-1-i\sqrt{2\la}\right)}}

\def\cUnoS{\frac{\hat a_1}\al+\hat b_1-\frac{i\pi\left(1-\e^{2ia\sqrt\la}\right)}{4\sqrt\la}}
\def\cDueS{\frac{\hat a_2}\al+\hat b_2+\frac{\pi\log(-\la)}4-\frac{\pi K_0(2a\sqrt{-\la})}2}
\def\cTreS{\frac{\hat a_3}\al+\hat b_3-\left(\frac{i\pi^2\sqrt\la}{2}-\frac{\pi^2\e^{2ia\sqrt\la}}{4a}\right)}

\def\trUnoS{\left(\frac{i\pi (1-\e^{2ia\sqrt\la})}{4\la^{3/2}}
   -\frac{\pi a\e^{2ia\sqrt\la}}{2\la}\right)
 \frac{1}{\cUnoS}}
\def\trDueS{\frac{\pi (1+2ia K_1(-2ia\sqrt\la))}{2\la} \frac{1}{\cDueS}}
\def\trTreS{\frac{\pi^2(1-\e^{2ia\sqrt\la})}{2i\sqrt\la} \frac{1}{\cTreS}}

\def\cngl{\frac{A_n}\alpha+\frac{g_n(i)+g_n(-i)}{4pi^3}-\frac{g_n(\la)}{2\pi^3}}

\section{Introduction}
\label{intro}

Recently, there has been a growing interest in the Casimir effect, namely the 
manifestation of vacuum energy at experimental as well as at theoretical
level (see, for example \cite{mostepanenko,milton} and references therein).
Because of the increasing 
interest in the Casimir effect and in spite of several results 
which have already been obtained, ``a solvable model''  
that permits  to obtain systematically explicit results is of 
the greatest interest. In this paper we present such a model. Moreover we put
at work mathematical techniques, which are of interest by themselves.

We shall study the Casimir effect 
related to a massless scalar field in a flat space-time modified by  
the presence of a pointlike (uncharged) ``impurity'',  
modelled by delta-like potentials, 
in manifolds with and without boundary. 
 
The boundaryless delta potential case has been
already treated (see for example \cite{ST,Par,Sol,irina,ku,unoA}. 
This case is also referred to as semi-transparent 
boundary conditions (see \cite{bordag,G,jaffe,jaffe1} and references therein). 
Here we shall deal with the new case of a delta potential on the 
half space.

On general grounds,  from one side, we present a rigorous 
mathematical description of the Schr\"{o}dinger-like  operators with 
delta-like potentials, and from the other side,  we will make use of 
a technique to regularize the 
functional determinant of self-adjoint elliptic operators 
defined on non compact manifolds associated with 
continuous spectrum. 

In order to start formulating the problem, we use the approach of  
Finite Temperature Quantum Field Theory based on the 
imaginary time formalism (see, for example \cite{gibb77} \cite{dowker78},  and \cite{ZZ,ZZ2,ZZ1} \cite{OS}, and references therein).  
We consider a massless scalar free field in 
four dimensional Minkowski space-time interacting with an external field 
represented by a potential $V$. Thus,  one is dealing with  the manifold 
$X(T)=S^1_{\beta/2\pi}\times M$, where $S^1_{r}$ is the circle of radius $r$,  
$\beta=\frac{1}{T}$, the period of the imaginary compactified time, is the inverse of 
the temperature, and $M$ is a three 
dimensional manifold.  The relevant   operator reads  
$H=-\Delta_{X(T)}+V=-\p_u^2-\Delta_M+V=-\p_u^2+L_M$, where $\Delta_Y$ is the 
Laplace-Beltrami 
operator on a manifold $Y$ defined by some Riemannian structure, 
and $V:M\to \R$ is a suitable potential. 

The canonical partition function at temperature $T$ of this model may be 
formally written as  
\beq
\log Z=-\frac{\beta}{2} \sum_{\lambda\in \Sp\,L_M }\lambda^{1/2}-
\sum_{\lambda\in \Sp \, L_M}\left(1-\e^{-\beta \sqrt{\lambda}}\right),
\label{LOGZ}
\eeq
here $\Sp\,L_M$ is the spectrum (a self-realization of) $L_M$.
We are assuming $M$ to be a compact manifold and $V$ a smooth potential. 
The first term on the right-hand side of equation (\ref{LOGZ}) 
corresponds to the vacuum energy contribution (Casimir energy), 
given by
\beq
E_c=-\lim_{\beta \rightarrow \infty}\partial_\beta \log Z=
\frac{1}{2} \sum_{\lambda\in \Sp \, L_M }\lambda^{1/2} ,
\label{EC}\eeq
while the second one, corresponding to the statistical sum contribution, is vanishing in the zero temperature limit. In order to give a meaning to the divergent first term in equation (\ref{LOGZ}), 
one may make use of the well-known zeta function regularization, namely one introduces the generalized zeta function, defined for large values of the real part of $s$ by  
\[
\zeta(s;L_M)=\sum_{\lambda\in \Sp \, L_M} \lambda^{-s},
\]
and by analytic continuation elsewhere, and one
 replaces equation (\ref{EC}) by $E_c=\frac{1}{2}\zeta(-1/2;L_M)$.

Nevertheless this approach does not work in general, because  
it may happen that $\zeta(s;L_M)$ is 
singular in $s=-1/2$. A possible approach is to consider $\log Z$ as a regularized functional determinant of the operator $H=-\p_u^2+L_M$, namely 
$\log Z=-\frac{1}{2}\zeta'(0,H)$ (see, for example, \cite{cognola92}). 
As a consequence, it is possible to show that it can be expressed in terms 
of some invariants of the geometric 
zeta function, i.e. the zeta function of the restriction of $H$ 
to $M$, and introducing another spectral function,
the  generalized Dedekind eta function \cite{OS}, 
defined for a positive operator $A$ with discrete spectrum by
\[
\eta(\tau;A)= \prod_{\lambda\in \sigma(A)}\left(1-\e^{-\tau\sqrt{\lambda}}\right).
\]

In fact, assuming that $-\Delta_M+V$ has trivial kernel, 
by Proposition 3 of \cite{OS} (see also  \cite{ZZ,ZZ2,cognola92}), we have 
\[
\zeta(0;H)=
\frac{1}{T}\Ru_{s=-\frac{1}{2}}\zeta(s;-\Delta_M+V),
\]
while by Corollary 1 of \cite{OS}
\[
\zeta'(0;H)=-\frac{1}{2T}\Rz_{s=-\frac{1}{2}} 
\zeta(s;-\Delta_M+V)-\frac{1-\log 2}{T}\Ru_{s=-\frac{1}{2}} \zeta(s;-\Delta_M+V)-2\log\eta\Big(\frac{1}{T};-\Delta_M+V\Big),
\]
where, for a meromorphic function $f(s)$,  $\Rz,\Rk$ are defined by means of the 
Laurent expansion
\[
f(s)=\Rz_{s=s_0}f(s)
+\sum_{k=1}^\infty\,\frac1{(s-s_0)^k}\,\Rk_{s=s_0}f(s).
\]

In this paper, we will consider explicitly differential operators with 
a singular potential on non compact manifolds with or without boundary.
Thus we shall need to generalize the above results to the non compact case. 
This will be done in details in the next sections. 

Now let us introduce the class of models we are going to investigate. 
In order to describe the class of operators we shall deal with, 
we start with an heuristic treatment which will be mathematically justified in 
the next sections. We first recall the Lipmann-Schwinger equations for an operator
$H=H_0+V$ defined in $L_2(\R^n)$, where $H_0=-\Delta$ is minus Laplacian and 
$V$ is a suitable non confining potential. They are given by
\[
\Psi^{\pm}( x)=\Psi_0^{\pm}( x)+\int_{\R^n} G^{(0)}_k( x,  y)V( y)  
\Psi^{\pm}( y)d  y ,\qquad x\in\R^n,
\]
where $ G^{(0)}_k( x,  y)$ is the Green function of the unperturbed 
operator $H_0$, namely
\[
(H_0-k^2) G^{(0)}_k(x,y)=\delta(x-y) .
\]

For example, for $n=1$, one has
\[
G^{(0)}_k( x,  y)=\frac{i}{2k}e^{ik|x-y|} .
\]

The above integral equation is the counterpart of the well known
resolvent identity associated with the resolvent of the operator $H$.

We shall consider  singular perturbations of the form  
\[
V( x)=g \delta( x-  a) ,
\]
where $g$ is the real coupling constant, and we limit our analysis to the cases
$n=1,2,3$, since only within these cases, one may implement delta-like interactions by self-adjoint operators in Hilbert space (see, for example \cite{AGHH}).  Hence, heuristically
\beq\label{introeq1}
H=-\Delta+g \delta(x-a).
\eeq

In this case, one formally has as solution
\[
\Psi^{\pm}( x)=\Psi_0^{\pm}( x)+ g G^{(0)}_k( x, a)  
\Psi^{\pm}( a) .
\label{ls1}
\]

Since $G^{(0)}_k( x, a) $ is singular when $ x \rightarrow  a$ for $n=2,\, 3$, 
the above solution of the original integral equation  is {\em inconsistent} 
and  one has to 
deal with a regularization and a renormalization procedure, first introduced in  \cite{BF}. 
Here we describe a regularization in the configuration space. First, the regularization 
may be achieved by making the replacement  $g \rightarrow g(\varepsilon)$ and
$G^{(0)}_k( x,  a) \rightarrow G^{(0)}_k(x+\varepsilon,a)$, for $y>0$.
As a result, neglecting terms which vanish as the cutoff $\epsilon$ is removed, 
i.e. when $\varepsilon\to0$, we may solve the above equation and arrive at 
\[
\Psi^{\pm}( a)=\Psi_0^{\pm}( a)+ g(\varepsilon)
 G^{(0)}_k( a, a+ \varepsilon) \Psi^{\pm}( a)\, .
\]

Thus, the regularized solution may be written as
\beq\label{introeq2}
\Psi^{\pm}( x)=\Psi_0^{\pm}( x)+ 
\frac{1}{\frac{1}{g(\varepsilon)}-
G^{(0)}_k( a,  a+ \varepsilon) }G^{(0)}_k( x, a)\Psi_0^{\pm}( a)\,   
\eeq

Furthermore, the renormalization consists in assuming that 
$g(\varepsilon)$ vanishes in the 
limit $\varepsilon\to0$ in such a way that
\[
\frac{1}{g(\varepsilon)}-G^{(0)}_k(a,a+\varepsilon)=
\frac{1}{g_R}-\Rz_{\varepsilon=0}G^{(0)}_k( a,a+\varepsilon)+O(\varepsilon),
\]
for some $g_R\neq0$.
As a consequence, one may remove the 
cutoff and one arrives at a finite expression, where renormalized quantities
appear, that is
\[
\Psi^{\pm}( x)=\Psi_0^{\pm}( x)+ 
\frac{1}{\frac{1}{g_R}-\Rz_{z=a}G^{(0)}_k(a,z)}\,G_k(x,a) \Psi_0^{\pm}( a).  
\]

For $n>3$, formally the above expression is still valid, but the interpretation
of $\Psi^{\pm}( x)$ as scattering states related to a self-adjoint Hamiltonian 
defined on an Hilbert space no longer holds. 

As an example, let us consider $n=3$ and $ a=0$. Then, one has
\[
G^{(0)}_k( x)=\frac{1}{4\pi | x|}e^{i k|x|} ,
\]
and 
\[
\Psi^{\pm}(x)=e^{\pm i kx}+ 
\frac{1}{\frac{1}{g_R}-i k}\frac{e^{i k|x|}}{|x|}.
\]

Instead, for $n=2$, one has
\[
G_k(x,y)=\frac{i}{4}H_0^{(1)}(k|x-y|),
\]
$H_0^{(1)}$ being a Hankel function.
Due to the presence of a logarithmic singularity for $x=y$, 
the regularization procedure leads to the appearance of an arbitrary dimensional 
scale $\ell $ and the regularized coupling constant has to be ``running'', 
in order to ensure the independence of the physical 
observables from $\ell$. The result is
\[
\Psi^{\pm}( x)=\Psi_0^{\pm}( x)+ 
\frac{(i/4) H_0^{(1)}(k|x|)}{\frac{1}{g_R(\ell)}+
\frac{1}{2\pi}(\ln{(k\ell/2i)}-\Psi(1)) } . 
\]

Coming back to the $n=3$ case, one may obtain the physical meaning of $g_R$, 
considering the non relativistic scattering of a particle of mass $m>0$. In this 
case, the operator $H_0$ is the kinetic energy 
(the Planck constant being taken to be one) 
and we have for the scattering 
wave-functions $ \Psi^{\pm}( x)$
\[
\Psi^{\pm}(\vec x)=e^{\pm i kx}+ 
\frac{2m g_R}{1-2im g_R k }\frac{e^{i k| x| }}{| x|} .
\]

The scattering amplitude can be read off and is
\[
f(k)=\frac{2m g_R}{1-2img_Rk} ,
\]
and the differential cross section is given by
\[
\frac{d \sigma}{d \Omega}=|f(k)|^2=\frac{4m^2 g_R^2}{1+4m^2g_R^2k^2}.
\]

The scattering length may be defined as
\[
a^2=\lim_{k \rightarrow 0}|f(k)|^2,
\]
in such a way that $\lim_{k \rightarrow 0} \sigma(k)=4\pi a^2$. Thus,
\[
a^2=4m^2g_R^2,
\]
namely the regularized coupling constant is proportional to the scattering 
length of the related non relativistic 3-dimensional scattering process.

It is easy to show that equation (\ref{introeq2}) is equivalent to the 
following expression for the kernels of the resolvents
\beq\label{introeq3}
G_\lambda(x,y)=G^{(0)}_\lambda(x,y)+
\frac{1}{\frac{1}{g_R}-\Rz_{z=a}G^{(0)}_k(a,z)}G^{(0)}_\lambda(x,a) 
 G^{(0)}_\lambda(y,a)\, .
\eeq

This formula is valid in general. For example, when the unperturbed 
operator $H_0$ is minus the Laplace operator defined in the
manifold with boundary $\R^+ \times \R^{n-1}$, 
we may repeat the above arguments and arrive at equation  (\ref{introeq3}), in
which $ G^{(0)}_\lambda( x, y)$ and $G_\lambda( x, y) $ now  satisfy a 
suitable boundary condition, for example the Dirichlet boundary condition
\[
G^{(0)}_\lambda( 0, y)=0 . 
\]

In the physically relevant case of $n=3$, we have
\[
G^{(0)}_\lambda( x,  y)=\frac{1}{4\pi} \left( \frac{e^{i 
\sqrt \lambda| x- y|}}{| x- y|  }-
 \frac{e^{i \sqrt \lambda| x-R  y|}}{| x-R y| } \right) ,
\]
where $R$ is the spatial reflection with respect to 
the plane which forms the boundary $\R^2$. For example, in this case, the 
renormalization leads to
\beq
\frac{1}{g(\varepsilon)}-G^{(0)}_\lambda( \varepsilon)=
\frac{1}{g_R}-\frac{\lambda}{4\pi}-  \frac{e^{-2\lambda a}}{2 a}+
O(\varepsilon)\, .
\label{rino}
\eeq

Again, for $n>3$, one may formally consider the above expressions, 
but without any references to some Hilbert space.  

In the following sections, making use of the method of self-adjoint extensions
and the general theory of singular perturbations, we will present a rigorous
mathematical derivation of the above heuristic results.

\section{Self adjoint extensions of non maximal multiplication operators}
\label{s1}

\subsection{General setting}
\label{s1.0}

Let $\H$ be an Hilbert space (complete and separable), and $A$ a self adjoint operator in $\H$. Fixing a suitable  restriction $\dot A$ of $A$, it is possible to construct a one parameter family of self adjoint operators $A_\alpha$ containing the initial operator $A$. 
This quite general setting was developed in \cite{AK}, 
Section 1.2.2. We recall here the main points of the construction, and we give a new proof of the main result, stated in  Lemma 1.2.3 of \cite{AK}. Let $|A|=(A^\dagger A)^\frac{1}{2}$, and for $s\geq 0$, let 
\[
\H_s=\D(|A|^\frac{s}{2})=\{v\in \H~|~ (|A|+I)^\frac{s}{2}v\in \H\}.
\]

Note that 
$\H_2=\D(A)$ and $\H_0=\H$. $\H_s$ is a complete Hilbert space with scalar product
\[
(u,v)_s=((|A|+I)^s u,v)_0.
\]

Obviously, $(|A|+zI)^\frac{s}{2}$ is an isometry of $\H_s$ onto $\H$ 
for all $s$ and for all $z$ that is neither zero nor a negative real number.
Let $\H_{-s}=\H_s^\dagger$ be the adjoint space of $\H_s$. We define the mapping 
\begin{align*}
(|\A|+zI)^{\frac{s}{2}}:&\H\to \H_{-s},\\
(|\A|+zI)^{\frac{s}{2}}:&u\mapsto (|\A|+zI)^{\frac{s}{2}}u,\\
\end{align*}
by 
\[
((|\A|+zI)^{\frac{s}{2}}u)(v)=(u,(|A|+zI)^\frac{s}{2} v)_0,
\]
for all $v\in \H_s$, and $\Re(z)> 0$. On the other hand, 
for each $\u\in \H_{-s}$, there exists a vector $u'\in \H_s$ such that
\[
\u(v)=(u',v)_s,
\]
for all $v\in \H_s$. It follows from the definition of $\H_s$, that $(|A|+zI)^\frac{s}{2}u'\in \H$, and 
\[
\u(v)=(u',v)_s=((|A|+zI)^\frac{s}{2}u',(|A|+zI)^\frac{s}{2}v)_0,
\]
for $\Re(z)> 0$. Therefore, we have a map
\begin{align*}
(|\A|+zI)^{-\frac{s}{2}}:&\H_{-s}\to \H,\\
(|\A|+zI)^{-\frac{s}{2}}:&\u\mapsto (|A|+zI)^\frac{s}{2}u'.\\
\end{align*}

It is easy to see that the mapping $(|\A|+zI)^{-\frac{s}{2}}$ 
and $(|A|+zI)^\frac{s}{2}$ are inverse to each other. If we define the scalar product 
\[
(\u,\v)_{-s}=((|\A|+zI)^{-\frac{s}{2}} \u,(|\A|+zI)^{-\frac{s}{2}} \v)_0,
\]
both maps preserve this scalar product. We have proved that $(|\A|+zI)^{s}$ is an 
isometry  of $\H_{t+2s}$ onto $\H_{t}$, for all real $s$ and $t$, 
and all $z$ with $\Re(z)> 0$, and  $|\A|=|A|$ when both $s,t\geq 0$. 
Beside these isometries, we have the obvious inclusion $\H_s\leq \H_t$, for all $s\geq t$. 
We also note that, using the scalar product in $\H_{-s}$, 
the action of the functional $\u\in\H_{-s}$ is
\[
\u(v)=(\u,(|A|+zI)^sv)_{-s},
\]
for all $v\in\H_s$, $\Re(z)> 0$.


\begin{lem} \label{l01} Let $A$ be a self adjoint operator in the Hilbert space $\H$, and $\ee\in \H_{-2}-\H$. Then, the restriction $\dot A$ of $A$ defined by the domain
\[
\D(\dot A)=\{v\in \H~|~Av\in \H, \ee(v)=0\},
\]
is symmetric, and has deficiency indices $(1,1)$. The solutions of the equation
\beq\label{eq0}
(\dot A^\dagger\mp iI)v=0,
\eeq
are all given by
\[
u_\pm = c (|\A|\mp iI)^{-1}\ee,
\]
with $c\in\C$.
\end{lem}

\noindent {Pooof.} Note that the domain is well defined. For $Av\in \H$ if and only if $v\in\D(A)$. First, we  prove that $\dot A$ is symmetric. We show that $\D(\dot A)^\perp=\{0\}$, where the orthogonal complement is in $\H$. In fact, if this is the case, then
\[
(\D(\dot A)^\perp)^\perp=\H,
\]
and $(\D(\dot A)^\perp)^\perp=\overline{\D(\dot A)}$, since $\D(\dot A)$ is a subspace, and the thesis follows.
By definition
\[
\D(\dot A)^\perp=\{v\in \H~|~(v,w)_0=0, \forall w\in \D(\dot A)\}.
\]

We show that a vector $v$ satisfies the equation 
\beq\label{eq1}
(v,w)_0=0,
\eeq
for all $w\in \D(\dot A)$, if and only if $v$ is a multiple of $\ee$ under the inclusion of $\H$ in $\H_{-2}$ (one implication is obvious, since $w\in \D(\dot A)$). This implies that the unique solution $v$ in $\H$ of the above equation is $v=0$.
We have the following facts:
\begin{itemize}
\item[(a)] if $w\in \D(\dot A)\leq \H_2$, then 
\[
(|A|+I)w=(|\A|+I)^{-1}(|\A|+I)^2w\in \H,
\]
and hence $(|\A|+I)^2 w\in \H_{-2}$;

\item[(b)] by definition, $w\in\D(\dot A)$ if and only if
\beq\label{eq2}
0=\ee(w)=(\ee,(|\A|+I)^2 w)_{-2};
\eeq

\item[(c)] $\D(\dot A)\leq \H_2$, and hence it is isometric to a subspace $D$ of $\H_{-2}$, $D=(|\A|+I)^2 \D(\dot A)$;

\item[(d)] each $v\in\H$ defines a functional $\v$ on $\H_2$ by $\v(w)=(v,w)_0$; thus, we rewrite equation (\ref{eq1}) as
\beq\label{eq3}
0=(v,w)_0=\v(w)=((|\A|+I)\v,(|\A|+I)w)_{-2}=(\v,(|\A|+I)^2w)_{-2},
\eeq
for all $w\in \D(\dot A)$, with $v=\v\in \H\leq \H_{-2}$.
\end{itemize}

By point (c), equation (\ref{eq2}) means that $D=L(\ee)^{\perp_{-2}}$, where $\perp_{-2}$ means that the orthogonal complement is in $\H_{-2}$. 
By point (d), $\v$ satisfies equation (\ref{eq1}) if and only if $\v\in D^{\perp_{-2}}$. 
Since $\H_{-2}=L(\ee)\oplus L(\ee)^{\perp_{-2}}$, it follows that $D^{\perp_{-2}}=(L(\ee)^{\perp_{-2}})^{\perp_{-2}}=L(\ee)$, and hence $\v$ satisfies equation (\ref{eq1}) if and only if $\v\in L(\ee)$, as required.

Next, we prove that the vectors $u_\pm$ are the unique solutions of the deficiency equation (\ref{eq0}). Note that since $\ee\in \H_{-2}$, it follows that $u_\pm\in\H$, and $\dot A u_\pm \notin \H$. We show that $u_\pm\in D(\dot A\dagger)$. 
By definition
\[
\D(\dot A^\dagger)=\{v\in\H~|~\exists u\in\H, (\dot A w,v)_0=(w,u)_0, \forall w\in \D(\dot A)\}.
\]

If we take $u=\pm iu_\pm$, then
\begin{align*}
(\dot A w,u_\pm)_0-(w,\pm iu_\pm)_0&=(\dot A w,u_\pm)_0-(\mp i w,u_\pm)_0=((\dot A\pm iI)w,u_\pm)_0\\
&=((A\pm iI) w,u_\pm)_0=(w,(A\mp iI)u_\pm)_0=\overline{\ee(w)}=0,
\end{align*}
since $A=\dot A$ on $\D(\dot A)$, and $A$ is self adjoint, and hence $u_\pm$ belong to $\D(\dot A^\dagger)$. This also means that $\dot A^\dagger u_\pm=i u_\pm$, and therefore the $u_\pm$ are solutions of equation (\ref{eq0}). It remains to show that these are the unique solutions. For, note that the solutions of equation (\ref{eq0}) are elements of the space
\[
\ker (iI-\dot A^\dagger)=(\Ran(-iI-\dot A))^\perp,
\]
and $u\in (\Ran(-iI-\dot A))^\perp$ if and only if
\beq\label{eq4}
(u,(-iI-\dot A)w)_0=0,
\eeq
for all $w\in\D(\dot A)$. By point (c) above, if $w\in \D(\dot A)$, then $(|\A|+I)^2w\in \H_{-2}$, so equation (\ref{eq4}) means that
\begin{align*}
0&=(u,(-iI-\dot A)w)_0=(u,(-iI-A)w)_0=(u,(-iI-A)(|\A|+I)^{-2}(|\A|+I)^2w)_0\\
&=((iI-\A)\u,(|\A|+I)^2w)_{-2},
\end{align*}
since $A$ is self adjoint. This implies that $(iI-\A)\u\in D^{\perp_{-2}}$ (where the space $D$ was defined in point (c) above). Since $D^{\perp_{-2}}=L(\ee)$, this completes the proof.

Using the standard von Neumann theory of self adjoint extensions, we characterize the adjoint and the self adjoint extensions of $\dot A$ as follows.

\begin{lem}\label{l3} The adjoint operator $\dot A^\dagger$ is
\begin{align*}
\D(\dot A^\dagger)&=\{w\oplus c_+u_+\oplus c_-u_-,w\in\D(\dot A),c_\pm\in\C\},\\
\dot A^\dagger(w+u_++u_-)&=\dot A w+iu_+-iu_-.
\end{align*}
\end{lem}

\begin{lem}\label{l4} All the self adjoint extensions $A_\theta$, $0\leq\theta<2\pi$, of the operator $\dot A$ are 
\begin{align*}
\D(A_\theta)&=\{w\oplus c_+(u_+\oplus\e^{i\theta}u_-),w\in\D(\dot A),c_+\in\C\},\\
A_\theta(w+u_++u_-)&=\dot A^\dagger(w+u_++u_-)=\dot Aw+iu_+-iu_-.
\end{align*}
\end{lem}

For proofs of these Lemmas see for example \cite{AK}, \cite{AG}. 
An equivalent description of the self adjoint extensions can be given by boundary conditions on the domain of the adjoint operator as
\[
\D(A_\theta)=\{u\in \D(\dot A^\dagger)~|~ (u_-,u)_0=\e^{i\theta} (u_+,u)_0\}.
\]

\begin{rem}\label{r1} Note that the case $\theta=\pi$ gives the maximal operator, namely $A_\theta\subseteq A_\pi$, for all $\theta$. For $v\in \D(Q_\pi)$ if and only if $v=w+c_+(u_+-u_-)$, with $w\in\D(\dot A)$, and $c_+\in\C$. But it is easy to see that the function $u=u_+-u_-=2i(\A^2+I)^{-1}\ee$ is such that $u\in L^2(\Omega^n)$ and $\dot A u\in L^2(\Omega^n)$. This means that $u\in \D(A)$, and the statement follows.
\end{rem}

Next we characterize the resolvent of the self adjoint extensions $A_\theta$ of $\dot A$. This should be compared with Theorem 1.2.1 of \cite{AK}.

\begin{lem}\label{l6} Let $A_\theta$ be one of the self adjoint extensions of the operator $\dot A$ described in Lemma \ref{l4}. Let $\lambda\in \rho(A_\theta)\cap \rho(A_\pi)$, then the resolvent of $A_\theta$ of $\dot A$ is
\[
R(\lambda, A_\theta)v=R(\lambda,A_\pi)v+c_\theta(\lambda)(u_{\bar\lambda},v)_0 u_\lambda, 
\]
where $c_\theta(\lambda)$ is some function of $\lambda$, and 
\[
u_\lambda=(|\A|-\lambda I)^{-1}\ee.
\]

Moreover, the difference of the resolvents $R(\lambda,A_\theta)-R(\lambda,A_\pi)$ is of trace class
, and
\[
\Tr(R(\lambda,A_\theta)-R(\lambda, A_\pi))=c_\theta(\bar\lambda)(u_{\bar\lambda},u_\lambda)_0.
\]
\end{lem}

\noindent {Pooof.} Let $\lambda\in \rho(A_\theta)\cap \rho(A_\pi)$. 
Consider 
\[
(R(\lambda,A_\theta)-R(\lambda,A_\pi))v=((\lambda I-A_\theta)^{-1}-(\lambda I-A_\pi)^{-1})v.
\]

Since for all $u\in \Ran (\lambda I-\dot A)$, we have that $(\lambda I-A_\theta)^{-1}u=(\lambda I-\dot A)^{-1} u$, it follows that 
\[
(R(\lambda,A_\theta)-R(\lambda,A_\pi))v=((\lambda I-A_\theta)^{-1}-(\lambda I-A_\pi)^{-1})P_{(\Ran(\lambda I-\dot A))^\perp}v,
\]
where $P_{(\Ran(\lambda I-\dot A))^\perp}$ denotes the projection onto $(\Ran(\lambda I-\dot A))^\perp$. 
But it is easy to see that the proof of Lemma \ref{l2} generalizes for any $\lambda\not=i$ in the resolvent set, thus $(\Ran(\lambda I-\dot A))^\perp=<u_{\bar\lambda}>$, with
\[
u_\lambda=(|\A|-\lambda I)^{-1}\ee.
\]

It follows that, 
\[
(R(\lambda,A_\theta)-R(\lambda,A_\pi))v=((\lambda I-A_\theta)^{-1}-(\lambda I-A_\pi)^{-1})(u_{\bar\lambda},v)_0 u_{\bar\lambda}.
\]

Now, the vector $y=((\lambda I-A_\theta)^{-1}-(\lambda I-A_\pi)^{-1}) u_{\bar\lambda}$ itself belongs to $(\Ran(\bar\lambda I-\dot A))^\perp$. For, since $\dot A^\dagger$ is an extension of $A_\theta$ for all $\theta$, 
\[
(\lambda I-\dot A^\dagger)y=(\lambda I-\dot A^\dagger)((\lambda I-A_\theta)^{-1}-(\lambda I-A_\pi)^{-1})u_{\bar\lambda}=0,
\]
implies that $y\in \ker (\lambda I-\dot A^\dagger)$.
Therefore, we have proved that for all $v\in \D(\lambda I-A_\theta)^{-1}$, 
\[
(R(\lambda,A_\theta)-R(\lambda,A_\pi))v=(u_{\bar\lambda},v)_0 c_\theta(\lambda)u_\lambda. 
\]

This means that 
\[
\Tr(R(\lambda,A_\theta)-R(\lambda,A_\pi))=c_\theta({\bar\lambda})(u_{\bar\lambda},u_\lambda)_0 <\infty,
\]
by the definition of $u_\lambda$, since $\ee\in\H_{-2}$.

\subsection{Multiplication operators} 
\label{s1.2} 

We pass now to consider a more concrete situation, namely multiplication operators. These operators provide the most natural setting where the results given in the previous section for abstract operators apply. Dually, all the results of the present section can be proved independently from the theory developed in Section \ref{s1.0}, but working directly in the concrete Sobolev spaces described below. We will not give complete proofs in this concrete setting, since they are precisely the same as the one provided in the abstract presentation of Section \ref{s1.0}. The main advantage working in this concrete setting, is that all the spaces $\H_s$ of Section \ref{s1.0} are subspaces of some large function space, and therefore all the functionals can be identified with some concrete functions in these spaces.
First, some preliminaries and notations. The measure appearing in all integrals is going to be 
Lebesgue's measure. 
Correspondingly, measurable sets and functions are understood in the sense of 
Lebesgue's integration theory.
Let $\Omega^n$ be some unbounded measurable subset of $\R^n$. Let $s$ be real, $q:\Omega^n\to \C$ be a measurable function, and
\[
m_s(x)=(1+|q(x)|)^\frac{s}{2}.
\]

We define the spaces
\[
L^{2,s}(\Omega^n)=\{f\in map(\Omega^n,\C)~|~fm_s\in L^2(\Omega^n)\}.
\]

Note that, $L^2(\Omega^n)=L^{2,0}(\Omega^n)$, and $L^{2,s}(\Omega^n)\subset L^{2,s'}(\Omega^n)$ if $s>s'$.  In $L^{2,s}(\Omega^n)$ we have the scalar product
\[
(f,g)_{L^{2,s}(\Omega^n)}=\int_{\Omega^n} \bar f(x) g(x) (m_s(x))^2 d^n x,
\]
and with this scalar product, the spaces $L^{2,s}(\Omega^n)$ are complete separable Hilbert spaces and are the Fourier images of the Sobolev spaces $W^{2,s}(\Omega^n)$.
We will use the notation $B^n_N$ for the intersection of $\Omega^n$ 
and the closed disc $D^n_N$ of radius $N$ centered in the origin of $\R^n$.

The maximal multiplication operator $Q$ associated to $q$ is the operator in $L^2(\Omega^n)$ defined by 
\begin{align*}
\D(Q)=&\{f\in L^2(\Omega^n)~|~ qf\in L^2(\Omega^n)\},\\
Qf=&qf.
\end{align*}

The operator $Q$ is a (closed) self adjoint operator with core 
$C_0^\infty (\Omega^n)$. If $q(x)\not=0$ a.e. in 
$\Omega^n$, then $Q$ is injective. 
If $q(x)\geq c$, for some $c$, a.e. in $\Omega^n$, 
then $\Ran(Q)=L^2(\Omega^n)$, so $Q:\D(Q)\to L^2(\Omega^n)$ is bijective.

\subsection{Non maximal multiplication operators and their extensions}
\label{s2}

Let $e:\Omega^n\to \C$ be a fixed measurable function. 
Assume the functions $q$ and $e$ decrease faster than some power, 
namely there exist constants $\mu$, $\beta$, $c$, and $c'$ such that
\[
|x^{-\mu} q(x)|\leq c,
\]
\[
|x^{-\beta} e(x)|\leq c'.
\]

Consider in $L^2(\Omega^n)$ the operator
\begin{align*}
\D(\dot Q)=&\left\{f\in L^2(\Omega^n)~|~ qf\in L^2(\Omega^n), \int_{\Omega^n} \bar e(x)f(x) d^n x=0\right\},\\
\dot Qf=&qf.
\end{align*}

It is clear that $\D(\dot Q)$ is a subspace of $L^2(\Omega^n)$, so the definition is well posed. Also, $\D(\dot Q)\subset \D(Q)$, so $\dot Q$ is a multiplication operator, but in general it is not maximal.

\begin{lem}\label{l2} If $\mu-\beta>\frac{n}{2}$, then $\dot Q$ is symmetric.
\end{lem}
\noindent {Pooof.} This follows from the first statement of Lemma \ref{l01}, provided that $e\in \H_{-2}$. In the present case, $|A|=|Q|$, and therefore $\H_{-2}=L^{2,-2}(\Omega^n)$. Thus, $e\in \H_{-2}$ if
\[
\int_{\Omega^n} \left|e(x)\right|^2(1+|q(x)|)^{-2} d^n x<\infty.
\]

We only need to check the convergence for large $r=|x|$. For large $r$ the integral behaves like $r^{2\beta-2\mu+n-1}$, and therefore it is convergent if $2\beta-2\mu+n<0$.

\begin{lem}\label{l3a} Assume $q$ is a real function, $\mu-\beta>\frac{n}{2}$, and $\beta\geq-\frac{n}{2}$. Then the operator $\dot Q$ has deficiency indices $(1,1)$, and the solutions of the equation
\[
(\dot Q^\dagger \pm iI)u=0,
\]
are all given by the functions
\[
\psi_\pm(x)=c\frac{e(x)}{q(x)\mp i},
\]
where $c\in \C,\, x\in\Omega^n$.
\end{lem}
\noindent {Pooof.} This follows from the second statement of Lemma \ref{l01}, provided that $e\in \H_{-2}$ and that $q\psi_\pm\notin L^2(\Omega^n)$. The first requirements implies $\mu-\beta>\frac{n}{2}$ as in the proof of the previous lemma. For the second one, consider the integral
\[
\int_{\Omega^n}|q(x)\psi_+(x)|^2d^n x=
\int_{\Omega^n}\frac{|q(x)e(x)|^2}{|q(x)+1|^2}d^n x.
\]

This integral behaves for large $r=|x|$ as $r^{2\beta+n}$, and therefore does not converge if $\beta\geq \frac{n}{2}$.

\begin{rem}\label{rpr} Note that the conditions $\mu-\beta>\frac{n}{2}$ and $\beta\geq -\frac{n}{2}$ imply that $\mu>0$. For $-\beta\leq \frac{n}{2}$, and hence $\frac{n}{2}<\mu-\beta\leq\mu+\frac{n}{2}$.
\end{rem}

We are now in the position of using the results in Lemmas \ref{l3} and \ref{l4} to characterize the adjoint of $\dot Q$ and to parameterize all the self adjoint extensions, 
using the parameter $\theta$. With this parameterization, 
the maximal multiplication operator $Q$ corresponds to the self adjoint extension defined by $\theta=\pi$ (see remark \ref{r1}). Using Lemma \ref{l6}, we also have a general formula for the resolvent and we know that the difference between the resolvents of a self adjoint extension and of the maximal operator is of trace class. 

We proceed by studying the particular case where $q(x)=|x|^2$ and $e$ is a bounded function. In this case, we give a more useful characterization of the self adjoint extensions of $\dot Q$ by some explicit integral boundary conditions. We will parameterize the self adjoint extensions 
by a real non negative parameter $\alpha$. 
By the assumptions on $q$ and $e$, we have $\mu=2$ and $\beta=0$, 
and the conditions in Lemmas \ref{l2} and \ref{l3a} are satisfied if 
and only if $n=1,2$ or $3$. Therefore, we proceed assuming $n$ 
to be in this range. In this case, if 
\[
\psi_\lambda(x)=\frac{e(x)}{|x|^2-\lambda}\, ,
\]
and assuming $\Im\la\neq0$,  
we have the following expansion for large $N$
\begin{equation}
\label{gnla}
\int_{B^n_N}\bar e(x)\psi_\la(x)d^n x=
   \int_{B^n_N}\frac{|e(x)|^2}{|x|^2-\la} d^n x=
   d_n(N)+g_n(\la)+\ot,\qquad n=1,2,3.
\end{equation}

Note that $d_n(N)$ does not depend on $\lambda$. The functions $d_n(N)$ and $g_n(\lambda)$ depend on the explicit form of $e(x)$. The values for the relevant choices of $e(x)$ are given in Lemma \ref{la1} in the appendix. In particular, it is always true that 
\[
d_1(N)=O\left(\frac1N\right),\qquad
d_2(N)=O\left(\log N\right),\qquad
d_3(N)=O\left(N\right).
\]

From equation (\ref{gnla}) it follows that
\begin{align*}
\int_{B^n_N}\bar e(x)\left(\psi_+(x)+\e^{i\theta}\psi_-(x)\right)d^nx
          &=(1+\e^{i\theta})d_n(N)+g_n(i)+\e^{i\theta}g_n(-i)+\ot\\
          &=i(1-\e^{i\theta})(\al d_n(N)+\al b_n+a_n)+\ot,
\end{align*}
where $\al,a_n,b_n$ are real numbers given by
\begin{align*}
\alpha&=\ctg\frac\theta2 ,\\
a_n&=\|\psi_+\|^2=\|\psi_-\|^2=
\int_{\Omega^n}\frac{|e(x)|^2}{1+x^4}d^nx=\frac{g_n(i)-g_n(-i)}{2i} ,\\
b_n&=\frac{g_n(i)+g_n(-i)}2 .
\end{align*}

Therefore, $\alpha$ is the real number which specifies the 
chosen self adjoint extension. 

This suggests the following description of the boundary conditions 
of the self adjoint extensions of $\dot Q$ in all the cases $n=1,2,3$.

\begin{prop}\label{l5} 
All the self adjoint extensions of the operator $\dot Q$ are given 
by the following integral boundary condition  
\begin{align*}
\D(Q_\alpha)&=\left\{f\in L^2(\Omega^n)~|~qf-\alpha c e\in L^2(\Omega^n), 
\phantom{\int_{B^n_N}}\right.\\ &\left.\qquad\qquad
\int_{B^n_N}\bar e(x)f(x)d^nx=c\left(\alpha d_n(N)+\alpha b_n+a_n\right)+o(1), ~for~large ~N
\right\},
\end{align*}
$c$ being an arbitrary complex constant, $d_n(N)$ the functions
given in equation (\ref{gnla}), and $\alpha$ a real number.
The operator $Q_\alpha$ acts on $f\in\D(Q_\alpha)$ as follows
\begin{align*}
(Q_\alpha f)(x)=&|x|^2 f(x)-\alpha c e(x)\\
=&|x|^2 f(x)-\frac{2\alpha}{\alpha-i}(\psi_+,f) e(x)\\
=&|x|^2 f(x)-\lim_{N\to\infty}\frac{\al e(x)}{\al d_n(N)+\al b_n+a_n}
\int_{B_N^n}\bar e(x)f(x)d^nx.
\end{align*}
\end{prop}

\begin{rem} \label{guido} In dimensions two and three, $d_n(N)$ are divergent quantities 
and so the constants $a_n,b_n$ can be dropped in the latter equation, while 
in dimension $n=1$ 
one can drop $d_1(N)$ and take the integral on the whole space.
\end{rem}

\noindent {Pooof.} Let $f\in\D(Q_\alpha)$, that is $f=h+c_+\left(\psi_++\e^{i\theta}\psi_-\right)$
with $h\in\D(\dot Q)$. Then,
\begin{align*}
\int_{B^n_N}\bar e(x)f(x)d^nx
   &=\int_{B^n_N}\bar e(x)h(x) d^nx
        +c_+ \int_{B^n_N}\bar e(x)
          \left(\psi_+(x) d^n x+\e^{i\theta}\psi_-(x)\right)d^nx\\
   &=\int_{B^n_N}\bar e(x)h(x) d^nx+c\left(\alpha d_n(N)+\alpha b_n+a_n\right)+\ot,
\end{align*}
with $c=i(1-\e^{i\theta})c_+$. Now we have to show that for large $N$, the integral in the latter equation always 
gives contributions which are negligible with respect 
$\alpha(d_n(N)+b_n)+a_n$. This is however a direct consequence of the fact that $h\in\D(\dot Q)$. 
For by hypothesis we have
\[
\int_{\Omega^n}\bar e(x)h(x)d^nx=0;
\]
since $\Omega^n=\bigcup_N B_N^n$, this implies that
\[
\lim_{N\to\infty}\int_{B^n_N}\bar e(x)h(x)d^nx=0,
\]
namely that 
\[
\int_{B^n_N}\bar e(x)h(x)d^nx=o(1),
\]
for large $N$.

The formula for the action of $Q_\alpha$ easily follows.

Next, we give the resolvent  of each extension.

\begin{prop}\label{ex1.l20} Let $Q_\alpha$ be one of the self adjoint extensions 
of the non maximal multiplication operator $\dot Q$ described in Proposition  
\ref{l5}. Then, for all $\lambda\in\rho(Q_\alpha)\cap\rho( Q_0)$, 
$Q_\alpha$ has the resolvent 
\begin{align*}
R(\lambda,Q_\alpha)\phi
   &=R(\lambda,Q_0)\phi+\frac{1}{\gna-g_n(\la)}
\int_{\Omega^n}\bar e(x)(R(\lambda,Q_0)\phi)(x)d^nx \psi_\lambda,\\
\psi_\lambda(x)&=\frac{e(x)}{|x|^2-\lambda},
\end{align*}
where the $g_n(\la)$ are the functions appearing in the asymptotic expansion of the integral boundary condition defining $Q_\alpha$ applied to the function $\psi_\lambda$, as given in equation (\ref{gnla}), and $Q_0$ is the maximal multiplication operator $Q$. Moreover, the difference of the resolvents $R(\lambda,Q_\alpha)-R(\lambda,Q_0)$ is of trace class.
\end{prop}

\noindent {Pooof.} By Lemma \ref{l6}, 
\[
R(\lambda,Q_\alpha)\phi=R(\lambda,Q_0)\phi+c_\alpha(\lambda,\phi) \psi_\lambda,
\]

In order to find the value of  $c_\alpha(\lambda,\phi)$,
first we note that $f=R(\lambda,Q_\alpha)\phi\in\D(Q_\alpha)$, 
therefore it must satisfies 
the conditions given in Proposition \ref{l5}, namely 
\begin{equation}
\label{ris1}
\int_{B^n_N}\bar e(x)f(x)d^nx
=\int_{B^n_N}\bar e(x)(R(\lambda,Q_\alpha)\phi)(x)d^nx
=c \left(a_n+\alpha b_n+\alpha d_n(N)\right)+o(1).
\end{equation}

On the other hand, using equation (\ref{gnla}) we explicitly have
\begin{align}
\label{ris2}
\int_{B^n_N}\bar e(x)(R(\lambda,Q_\alpha)\phi)(x)d^nx
    =&\int_{ B^n_N}\bar e(x)(R(\lambda,Q_0)\phi)(x)d^nx
  \\
     &+c_\alpha(\la,\phi) \int_{B^n_N}\bar e(x)\psi_\la(x)d^nx
\nonumber\\
    =&\int_{B^n_N}\bar e(x)(R(\lambda,Q_0)\phi)(x)d^nx
\nonumber\\
    &+c_\alpha(\la,\phi) \left(d_n(N)+g_n(\la)\right)+\ot.
\nonumber\end{align}

Now, since $R(\lambda,Q_0)\phi\in {\rm dom} (Q_0)$, by Proposition \ref{l5}
\[
\int_{B^n_N}\bar e(x)(R(\lambda,Q_0)\phi)(x)d^nx
=c a_n+o(1),
\]
for large $N$. Note that the constant $a_n$ does not depend on the extension by definition.

This means that we are able to make the comparison between the two equations (\ref{ris1})
and (\ref{ris2}). We have
\[
\int_{\Omega^n}\bar e(x)(R(\lambda,Q_0)\phi)(x)d^nx
+c_\alpha(\la,\phi) \left(d_n(N)+g_n(\la)\right)
-c(a_n+\alpha b_n+ \alpha d_n(N))=o(1),
\]
and this implies that $
c_\alpha(\lambda,\phi)=\alpha c
$
and
\[
c_\alpha(\lambda,\phi)=\frac{\alpha}{a_n+\alpha b_n-\alpha g_n(\la)}
\int_{\Omega^n}\bar e(x)(R(\lambda,Q_0)\phi)(x)d^nx
=\frac{1}{\gna-g_n(\la)}
\int_{\Omega^n}\frac{\bar e(x)\phi(x)}{\la-|x|^2}d^nx.
\]

Note that in the case $n=1$, we are comparing constants, since $d_1(N)$ is small in this case (see Remark \ref{guido}).

In order to give the kernel of the resolvent, we need a suitable delta function in the space $\Omega^n$. This will be denoted by $\delta_{\Omega^n}$, and is defined by the property
\[
\int_{\Omega^n}\delta_{\Omega^n}(x-a)f(x)d^nx=f(a),
\]
in the appropriate space of test functions over $\R^n$. 
Explicit formulas will be given in the concrete examples studied below.

\begin{corol} \label{cc10} 
The operator $R(\lambda,Q_\alpha)$ of Proposition \ref{ex1.l20} 
is an integral operator with kernel
\[
\ker(x,y;R(\lambda,Q_\alpha))=\frac{\delta_{\Omega^n}(x-y)}{\lambda-|x|^2}
                 -\frac{1}{\gna-g_n(\lambda)} 
                 \frac{e(x) \bar e(y)}{(\la-|x|^2) (\la-|y|^2)}.
\]
\end{corol}
\noindent {Pooof.} 
Since 
\[
\left(R(\lambda,Q_0)\phi\right)(x)=\frac{\phi(x)}{\lambda-|x|^2},
\]
we have that the operator
\begin{align*}
(A_\lambda\phi)(x)
    &=\psi_\lambda(x) \int_{\Omega^n}\bar e(y)(R(\lambda,Q_0)\phi)(y)d^ny\\
    &=\psi_\lambda(x) \int_{\Omega^n}\bar e(y)\frac{\phi(y)}{\lambda-|y|^2}d^ny
    =\int_{\Omega^n}\frac{e(x) \bar e(y)}{(|x|^2-\lambda) (\la-|y|^2)}\phi(y)d^ny,
\end{align*}
is an integral operator with kernel
\[
\ker (x,y;A(\lambda))=-\frac{e(x)}{\la-|x|^2}   \frac{\bar e(y)}{\la-|y|^2}.
\]

\begin{corol} \label{cc11} 
The difference of the resolvents $R(\lambda,Q_\alpha)-R(\lambda,Q_0)$  of Proposition \ref{ex1.l20} 
is a trace class operator with trace
\[
\Tr(R(\lambda,Q_\alpha)-R(\lambda,Q_0))=
-\frac{1}{\gna-g_n(\la)}\int_{\Omega^n}\frac{|e(x)|^2}{(|x|^2-\la)^2} d^nx.
\]
\end{corol}

\begin{rem}\label{l9.10} Assuming  $\Im \lambda\neq0$, we have the bound
\[
\left|\Tr(R(\lambda,Q_\alpha)-R(\lambda,Q_0))\right|
    \leq K \frac{1}{|\gna-g_n(\la)|} |\la^{\frac{n}2-2}|,
\]

with some positive constant $K$.  For by Corollary \ref{cc10} 
\[
\ker (x,y;R(\lambda,Q_\alpha)-R(\lambda,Q_0))
=-\frac{1}{\gna-g_n(\lambda)} 
                 \frac{e(x) \bar e(y)}{(\la-|x|^2) (\la-|y|^2)}.
\]

Since $\Im \la\neq0$ we can integrate obtaining
\begin{align*}
\left|\Tr(R(\lambda,Q_\alpha)-R(\lambda,Q_0))\right|
   &=\frac{1}{|\gna-g_n(\lambda)|} 
       \int_{\Omega^n}\frac{|e(x)|^2}{(|x|^2-\la)^2} d^nx\\
   &\leq\frac{1}{|\gna-g_n(\lambda)|} 
       \int_{\R^n}\frac{|e(x)|^2}{(|x|^2-\la)^2} d^nx\\
   &\leq  \frac{K}{|\gna-g_n(\la)|} |\la^{\frac{n}2-2}|.
\end{align*}

In the latter line we have used Lemma \ref{aa1} in the appendix.
\end{rem}

We conclude this section giving the continuum spectrum of the operators $Q_\alpha$. 
Possible isolated eigenvalues of finite multiplicity will be detected by an 
explicit study of the resolvent of particular examples in the following sections.

\begin{lem}\label{l7} The continuum spectrum of the operator $Q_\alpha$ coincides with the non negative real axis, i.e. $\Sp_c Q_\alpha=[0,\infty)$.
\end{lem}
\noindent {Pooof.} We recall that $\Sp_c\subseteq \Sp_e$, for closed operators, and $\Sp_e=\overline{\Sp_c}\cup\overline{\Sp_p}\cup\Sp_p^\infty$, for self adjoint operators. Then, the thesis follows since all self adjoint extensions have the same essential spectrum, and the maximal operator $Q_0$ is known to have the pure continuous spectrum $\Sp Q_0=\Sp_c Q_0=[0,\infty)$.

\subsection{The basic example in the whole space}\label{ex1} 
Let $\Omega^n=\R^n$, $e(x)=\e^{iax}$, where $a\in\R^n$, and $n=1,2,3$. 
We study in this section  the self adjoint extensions of the (closure of the) 
operator $\dot Q$ in $L^2(\R^n)$ defined by
\begin{align*}
\D(\dot Q)=&\left\{f\in L^2(\R^n)~|~ qf\in L^2(\R^n), 
     \int_{\R^n} \e^{-iax}f(x) d^n x=0\right\},\\
 \dot Q f=&qf,
\end{align*}
where $q(x)=|x|^2$. We have $|e|=1$, and $B_N^n=D_N^n$.
In order to apply Proposition \ref{l5}, we have to compute
the functions $d_n(N)$ and $g_n(\la)$ appearing in equation (\ref{gnla}).
They are explicitly given in Lemma \ref{la1} in the appendix. 
 
In the following we shall assume $\Im \sqrt\la>0$, then,
from  Lemma \ref{la1} in the appendix, for $n=3,2,1$ we have respectively 
\[\begin{array}{l}
n=3\,,\qquad\left\{\begin{array}{l}
\int_{D^3_N} \frac{1}{|x|^2-\lambda}d^3x=4\pi N+2\pi^2i \rl+o(1),\\
d_3(N)=4\pi N,\quad g_3(\la)=2\pi^2i \sqrt\la,
   \qquad a_3=\sqrt2 \pi^2 ,\quad b_3=-\sqrt2 \pi^2 ,
\end{array}\right.
\\\\
n=2\,,\qquad\left\{\begin{array}{l}
\int_{D^2_N} \frac{1}{|x|^2-\lambda}d^2x=2\pi\log N-\pi\log(-\lambda)+o(1),\\
d_2(N)=2\pi\log N,\quad g_2(\la)=-\pi\log(-\la), 
   \qquad a_2=\frac{\pi^2}{2}, \quad b_2=0 ,
\end{array}\right.
\\\\
n=1\,,\qquad\left\{\begin{array}{l}
\int_{D^1_N} \frac{1}{|x|^2-\lambda}dx=\frac{i\pi}{\rl}-\frac{2}{N}+o\left(\frac{1}{N}\right),\\
d_1(N)=-\frac2N,\quad g_1(\la)=\frac{i\pi}{\sqrt\la} ,
   \qquad a_1=\frac{\pi}{\sqrt2} ,\quad b_1=\frac{\pi}{\sqrt2} .
\end{array}\right.
\end{array}\]

Applying Propositions \ref{l5} and \ref{ex1.l20}, and Corollary \ref{cc11}, we easily prove  the following results.

\begin{lem}\label{ex1.l1} 
The self adjoint extensions of $\dot Q$ are:
\[\begin{array}{l}
n=3\,,\qquad\left\{\begin{array}{l}
\D(Q_\alpha)=\left\{f\in L^2(\R^3)~|~Q_\alpha f\in L^2(\R^3),
\phantom{\int_{D^3_N}}\right.\\ \left.\qquad\qquad\qquad 
\int_{D^3_N} \e^{-iax}f(x)d^3 x=c(4\pi N\alpha-\sqrt2\pi^2\alpha+\sqrt2\pi^2)
+o(1),c\in\C\right\},\\
(Q_\alpha f)(x)=|x|^2f(x)-\lim_{N\to\infty}\frac{\e^{iax}}
     {4\pi N}\int_{D^3_N}\e^{-iax}f(x)d^3 x,
\end{array}\right.
\\\\
n=2\,,\qquad\left\{\begin{array}{l}
\D(Q_\alpha)=\left\{f\in L^2(\R^2)~|~Q_\alpha f\in L^2(\R^2),
\phantom{\int_{D^2_N}}\right.\\ \left.\qquad\qquad\qquad 
\int_{D^2_N} \e^{-iax}f(x)d^2x=c\left(2\pi\alpha\log N
                  +\frac{\pi^2}2\right)+o(1),c\in\C\right\},\\
(Q_\alpha f)(x)=|x|^2f(x)-\lim_{N\to\infty}
\frac{\e^{iax}}{2\pi\log N}
\int_{B^2_N}\e^{-iax}f(x)d^2x,
\end{array}\right.
\\\\
n=1\,,\qquad\left\{\begin{array}{l}
\D(Q_\alpha)=\left\{f\in L^2(\R)~|~Q_\alpha f\in L^2(\R),
\phantom{\int_{D^1_N}}\right.\\ \left.\qquad\qquad\qquad  
    \int_{D^1_N} \e^{-iax}f(x)dx=c\left(-\frac{2\alpha}{N}
         +\frac{\pi\alpha}{\sqrt2}+\frac{\pi}{\sqrt2}\right)+o(1/N),c\in\C\right\},\\
(Q_\alpha f)(x)=|x|^2f(x)-\frac{\sqrt2\alpha \e^{iax}}{\pi(\al+1)}
\int_{\R}\e^{-iax}f(x)dx.
\end{array}\right.
\end{array}\]
\end{lem}

\begin{lem}\label{ex1.l2} 
Let $Q_\alpha$ be any of the self adjoint extensions of the non maximal multiplication 
operator $\dot Q$ described in Lemma \ref{ex1.l1}. 
Then, for all $\lambda\in\rho(Q_\alpha)\cap\rho(Q_0)$, 
$Q_\alpha$ has the resolvent
\[\begin{array}{ll}
n=3\,,&\qquad
R(\lambda,Q_\alpha)\phi
   =R(\lambda,Q_0)\phi+\frac{1}{\cTre} 
\int_{\R^3}\e^{-iax}(R(\lambda,Q_0)\phi)(x)d^3x \psi_\lambda,
\\
n=2\,,&\qquad
R(\lambda,Q_\alpha)\phi
   =R(\lambda,Q_0)\phi+\frac{1}{\cDue} 
\int_{\R^2}\e^{-iax}(R(\lambda,Q_0)\phi)(x)d^2x \psi_\lambda,
\\
n=1\,,&\qquad
R(\lambda,Q_\alpha)\phi
   =R(\lambda,Q_0)\phi+\frac{1}{\cUno} 
\int_{\R}\e^{-iax}(R(\lambda,Q_0)\phi)(x)dx \psi_\lambda,
\end{array}\]
where $\psi_\lambda(x)=\frac{\e^{iax}}{|x|^2-\lambda}$ for all the cases.

\end{lem}

\begin{corol} \label{cc1} 
The operator $R(\lambda,Q_\alpha)$ of Lemma \ref{ex1.l2} 
is an integral operator with kernel
\[\begin{array}{ll}
n=3\,,&\qquad
\ker(x,y;R(\lambda,Q_\alpha))=\frac{\delta(x-y)}{\lambda-|x|^2}
                 -\frac{1}{\cTre} 
                 \frac{\e^{ia(x-y)}}{(\la-|x|^2) (\la-|y|^2)},
\\
n=2\,,&\qquad
\ker(x,y;R(\lambda,Q_\alpha))=\frac{\delta(x-y)}{\lambda-|x|^2}
                 -\frac{1}{\cDue} 
                 \frac{\e^{ia(x-y)}}{(\la-|x|^2) (\la-|y|^2)},
\\
n=1\,,&\qquad
\ker(x,y;R(\lambda,Q_\alpha))=\frac{\delta(x-y)}{\lambda-|x|^2}
                 -\frac{1}{\cUno} 
                 \frac{\e^{ia(x-y)}}{(\la-|x|^2) (\la-|y|^2)},
\end{array}
\]
where $\lambda\in\rho(Q_\alpha)\cap\rho(Q_0)$ and $x,y\in\R^n$, $n=3,2,1$ 
respectively.
\end{corol}

Moreover, we have:

\begin{lem}\label{l9.1} 
Let $Q_\alpha$ be any of the self adjoint extensions of the non maximal 
multiplication operator $\dot Q$ described in Lemma \ref{ex1.l1}, then
the difference $R(\la,Q_\alpha)-R(\la,Q_0)$ is a trace-class operator
and, assuming $\Im \sqrt\la>0$

\[\begin{array}{ll}
n=3\,,&\qquad
\Tr(R(\lambda,Q_\alpha)-R(\lambda,Q_0))=\trTre ,
\\
n=2\,,&\qquad
\Tr(R(\lambda,Q_\alpha)-R(\lambda,Q_0))=\trDue ,
\\
n=1\,,&\qquad
\Tr(R(\lambda,Q_\alpha)-R(\lambda,Q_0))=\trUno .
\end{array}
\]
\end{lem}

\noindent {Pooof.} One has to compute the $L^2$-trace using the kernel given in 
Corollary \ref{cc1} and the results of Lemma \ref{aa1} in the appendix.


\subsection{The basic example in the half space} \label{ex2} 
Let $\Omega^n=\HH^n=[0,\infty)\times \R^{n-1}$, 
$e(x)=\sin(ax)$, 
where $a\in \HH^n$, and  $n=1,2,3$. 
We study in this section the self adjoint extensions of the 
(closure of the) operator $\dot Q$ in $L^2(\HH^n)$ defined by
\begin{align*}
\D(\dot Q)=&\left\{f\in L^2(\HH^n)~|~qf\in L^2(\HH^n),f(\{0\}\times \R^{n-1})=0, 
\int_{\HH^n} \sin(ax)f(x) d^nx=0\right\},\\
\dot Qf=&qf,
\end{align*}
where $q(x)=|x|^2$. We have $|e|=1$, 
and $B_N^n=D_N^n\cap \HH^n$, and the functions 
$d_n(N)$ and $g_n(\la)$ appearing in equation (\ref{gnla}) 
are given in Lemma \ref{la1b} in the appendix.

Recalling that the space of the functions satisfying Dirichlet boundary condition on the boundary of the half space naturally identifies with the space of the odd functions on the whole space, we realize the delta function in the half space as 
\[
\delta_{\HH^n}(x-a)=\frac{1}{2}(\delta(x-a) -\delta (x-P_n(a)),
\]
where $P_n$ is the reflection on the last coordinate.



Now we are able to write down explicitly all relevant quantities concernig the operator
$\dot Q$ in the half-space $\HH^n$ for $n=3,2,1$.

Assuming $\Im \sqrt\la>0$, from Lemma \ref{la1b} in the appendix we have
\[\begin{array}{l}
n=3\,,\qquad\left\{\begin{array}{l}
\int_{B^3_N}\frac{\sin^2(ax)}{|x|^2-\lambda}d^3x
   =\pi N+\frac{i\pi^2}{2}\rl-\frac{\pi^2}{4a} \e^{2ia\rl}+o(1),\\
   \hat d_3(N)=\pi N,
    \qquad\hat g_3(\la)=\frac{i\pi^2}{2}\sqrt\la-\frac{\pi^2}{4a} \e^{2ia\sqrt\la} ,     
     \\
      \hat a_3=\frac{\pi^2}{4}\left(\sqrt2+\frac{\e^{-\sqrt2a}\sin(\sqrt2a)}{2a}\right),
      \qquad\hat b_3=-\frac{\pi^2}{4}\left(\sqrt2+\frac{\e^{-\sqrt2a}\cos(\sqrt2a)}{2a}\right),
\end{array}\right.
\\\\
n=2\,,\qquad\left\{\begin{array}{l}
\int_{B^2_N}\frac{1}{|x|^2-\lambda}d^2x=\frac\pi2 \log N
   -\frac\pi4 \log(-\lambda)+\frac\pi2 K_0(2a\sqrt{-\lambda})+o(1),\\
   \hat d_2(N)=\frac\pi2 \log N,
    \qquad\hat g_2(\la)=-\frac\pi4 \log(-\lambda)+\frac\pi2 K_0(2a\sqrt{-\lambda}),
     \\
      \hat a_2=\frac\pi4\left(\frac\pi2+iK_0(2a\sqrt{i})-iK_0(2a\sqrt{-i})\right) ,
       \qquad\hat b_2=-\frac{i\pi}4\left(K_0(2a\sqrt{i})+K_0(2a\sqrt{-i})\right) ,
\end{array}\right.
\\\\
n=1\,,\qquad\left\{\begin{array}{l}
  \int_{B^1_N} \frac{1}{|x|^2-\lambda}dx
    =\frac{i\pi}{4\rl}\left(1-\e^{2ia\rl}\right)
       -\frac{1}{2N}+o\left(\frac{1}{N}\right),\\
   \hat d_1(N)=-\frac2N,
    \qquad\hat g_1(\la)=\frac{i\pi\left(1-\e^{2ia\sqrt\la}\right)}{4\sqrt\la},\\
     \hat a_1=\frac{\pi}{4\sqrt2}\left(1-\e^{-\sqrt2a}(\cos(\sqrt2a)+\sin(\sqrt2a))\right) ,
    \qquad\hat b_1=\frac{\pi}{4\sqrt2}\left(1-\e^{-\sqrt2a}(\cos(\sqrt2a)-\sin(\sqrt2a))\right) ,
\end{array}\right.
\end{array}\]
$K_0(z)$ being a Bessel function. Here we use the ``hat'' to distinguish the latter quantities
with respect to the ones appearing in the example discussed in Section \ref{ex1}, 
concerning the case of the  whole space.
 
Now, applying Propositions \ref{l5} and \ref{ex1.l20}, 
and Corollary \ref{cc11}, we prove the following results.
\begin{lem}\label{ex2.l1} The self adjoint extensions of $\dot Q$ are:
\[\begin{array}{l}
n=3\,,\qquad\left\{\begin{array}{l}
\D(Q_\alpha)=\left\{f\in L^2(\R^3)~|~Q_\alpha f\in L^2(\R^3),f(\{0\}\times \R^2)=0,
\phantom{\int_{B^3_N}}\right.\\ \left.\qquad\qquad\qquad
\int_{B^3_N}\sin(ax)f(x)d^3x
    =c(\pi N\alpha+\alpha\hat b_3+\hat a_3)+o(1),c\in\C\right\},\\
    Q_\alpha f=|x|^2f(x)-\lim_{N\to\infty} 
       \frac{\sin(ax)}{\pi N}\int_{B^3_N}\sin(ax)f(x)d^3x,
\end{array}\right.
\\\\
n=2\,,\qquad\left\{\begin{array}{l}
   \D(Q_\alpha)=\left\{f\in L^2(\HH^2)~|~Q_\alpha f\in L^2(\HH^2), f(\{0\}\times \R)=0,
   \phantom{\int_{B^2_N}}\right.\\ \left.\qquad\qquad\qquad
    \int_{B^2_N}\sin(ax)f(x)d^2x=c\left(\frac{\pi\alpha\log N}2+\alpha\hat b_2+\hat a_2\right)
    +o(1),c\in\C\right\},\\
   (Q_\alpha f)(x)=|x|^2f(x)-\lim_{N\to\infty}
  \frac{2\sin(ax)}{\pi\log N}\int_{B^2_N}\sin(ax)f(x)d^2x,
\end{array}\right.
\\\\
n=1\,,\qquad\left\{\begin{array}{l}
\D(Q_\alpha)=\left\{f\in L^2(R)~|~Q_\alpha f\in L^2(R), f(0)=0,
\phantom{\int_{B^1_N}}\right.\\ \left.\qquad\qquad\qquad
\int_{B^1_N} \sin(ax)f(x)dx=c\left(-\frac{2\alpha}N+\alpha\hat b_1+\hat a_1\right)
+o(1/N),c\in\C\right\},\\
(Q_\alpha f)(x)=|x|^2f(x)-\frac{4\sqrt2\alpha \sin(ax)\int_{\HH^1}\sin(ax)f(x)dx}
{\pi(1+\al)(1-\e^{-\sqrt2 a}\cos(\sqrt2 a))-\pi(1-\al)\e^{-\sqrt2 a}\sin(\sqrt2 a)
}.
\end{array}\right.
\end{array}\]

\end{lem}

\begin{lem}\label{ex2.l2} 
Let $Q_\alpha$ be any of the self adjoint extensions 
of the non maximal multiplication 
operator $\dot Q$ described in Lemma \ref{ex2.l1}. 
Then, for all $\lambda\in\rho(Q_\alpha)\cap\rho(Q_0)$, 
$Q_\alpha$ has the resolvent
(we assume $\Im\sqrt\la$ to be positive)
\[\begin{array}{ll}
n=3\,,&\qquad
   R(\lambda,Q_\alpha)\phi=R(\lambda,Q_0)\phi
     +\frac{\int_{\HH^3}\sin(ax)(R(\lambda,Q_0)\phi)(x)d^3x}{\cTreS}\psi_\lambda,
\\\\
n=2\,,&\qquad
   R(\lambda,Q_\alpha)\phi=R(\lambda,Q_0)\phi
     +\frac{\int_{\HH^2}\sin(ax)(R(\lambda,Q_0)\phi)(x)d^2x}{\cDueS}\psi_\lambda,
\\\\
n=1\,,&\qquad
   R(\lambda,Q_\alpha)\phi=R(\lambda,Q_0)\phi+
     \frac{\int_{\HH^1}\sin(ax)(R(\lambda,Q_0)\phi)(x)dx}{\cUnoS}\psi_\lambda,
\end{array}\]
where $\psi_\lambda(x)=\frac{\sin(ax)}{|x|^2-\lambda}$ for all the cases.
\end{lem}

\begin{corol} 
The operator $R(\lambda,Q_\alpha)$ of Lemma \ref{ex2.l2} 
is an integral operator with kernel 
\[\begin{array}{ll}
n=3\,,&\qquad
\ker(x,y;R(\lambda,Q_\alpha))=\frac{\delta_{\HH^3}(x-y)}{\lambda-|x|^2}-\frac{1}{\cTreS} 
   \frac{\sin(ax)\sin(ay)}{(\la-|x|^2) (\la-|y|^2)},
\\\\
n=2\,,&\qquad
\ker(x,y;R(\lambda,Q_\alpha))=\frac{\delta_{\HH^2}(x-y)}{\lambda-|x|^2}-\frac{1}{\cDueS} 
   \frac{\sin(ax)\sin(ay)}{(\la-|x|^2) (\la-|y|^2)},
\\\\
n=1\,,&\qquad
\ker(x,y;R(\lambda,Q_\alpha))=\frac{\delta_{\HH^1}(x-y)}{\lambda-|x|^2}-\frac{1}{\cUnoS}
   \frac{\sin(ax)\sin(ay)}{(\la-|x|^2) (\la-|y|^2)},
\end{array}\]
where $\lambda\in\rho(Q_\alpha)\cap\rho(Q_0)$, $x,y\in \HH^n$, $n=3,2,1$.
\end{corol}

Moreover, we have:

\begin{lem}\label{l9.1b} 
Let $Q_\alpha$ be any of the self adjoint extensions of the non maximal 
multiplication operator $\dot Q$ described in Lemma \ref{ex1.l1}, then
the difference $R(\la,Q_\alpha)-R(\la,Q_0)$ is a trace-class operator
and, assuming $\Im \sqrt\la>0$  
\[\begin{array}{ll}
n=3\,,&\qquad
\Tr(R(\lambda,Q_\alpha)-R(\lambda,Q_0))=\trTreS ,
\\\\
n=2\,,&\qquad
\Tr(R(\lambda,Q_\alpha)-R(\lambda,Q_0))=\trDueS ,
\\\\
n=1\,,&\qquad
\Tr(R(\lambda,Q_\alpha)-R(\lambda,Q_0))=\trUnoS ,
\end{array}\]
where $K_1$ is a Bessel function.
\end{lem}
\noindent {Pooof.} All latter integrals are computed in Lemma \ref{aa2} in the appendix.


\section{The Laplace operator with delta potential}
\label{lap}

We show in this section how the extensions of the multiplication operators introduced in Section \ref{s2} are used in order to define a self adjoint differential operator corresponding to the formal Laplacian operator with a delta type potential, discussed in Section \ref{intro}. This was the original approach of Berezin and Fadeev \cite{BF}. Let 
\[
-d=-\sum_{j=1}^n\frac{\partial^2}{\partial x_j^2},
\]
denotes the formal Laplace operator on $\Omega^n$, where $\Omega^n$ is either $\R^n$ or $\HH^n$. Let $a$ be a point in $\Omega^n$ and $\F$ denotes the Fourier transform in $\Omega^n$. Then, the operator $-\dot\Delta=\F^{-1} \dot Q \F$, is a (closed) symmetric operator in $L^2(\Omega^n)$, with deficiency indices $(1,1)$, for $n\leq 3$,  and domain
\[
\D (-\dot\Delta)=\{f\in W^{2,2}(\Omega^n)~|~f(a)=0\},\qquad
-\dot\Delta f=-d f,
\]
unitary equivalent to the operator $\dot Q$ defined at the beginning of Section \ref{s2}, with $e(x)$ either $\e^{iax}$ when $\Omega^n=\R^n$ or $\sin(ax)$ when $\Omega^n=\HH^n$. 
This follows immediately in both cases by the definition of the Fourier transform.
Therefore, all the self adjoint extensions of $-\dot\Delta$ are the operators $-\Delta_\alpha=\F^{-1} Q_\alpha \F$, where the operators $Q_\alpha$ were defined in general in 
Proposition \ref{l5}, and in the particular case of dimensions $n=3,2$ and $1$ 
in Lemmas \ref{ex1.l1} and  \ref{ex2.l1} respectively when $\Omega^n=\R^n$ or $R^n$. 
In all cases, the maximal operator is $-\Delta_0$ with
\[
\D (-\Delta_0)=W^{2,2},\qquad
-\Delta_0 f=-d f,
\]

It is worth to observe here that the operator $\dot\Delta$ can also be introduced directly (this is the approach of \cite{AGHH}) as the closure of the operator
\[
\D (-\tilde\Delta)=C_0^\infty (\Omega^n-\{a\}),\qquad
-\tilde\Delta f=-d f.
\]

This follows adapting the standard proof for the maximal operators (see for example \cite{Wei} 10.11). For it is clear that if $f\in\D(-\dot\Delta)$, then $f(a)=(\F^{-1}\F f)(a)=0$, and therefore $f\in\D(\dot Q)$. Conversely, given $f\in\D(\dot Q)$, we have that $(\F^{-1}f)(a)=0$. Since the closure of the restriction of $\dot Q$ on $C^\infty_0(\Omega^n)$ is closed, given  a sequence $\{h_n\}$ in $C^\infty_0(\Omega^n)$ such that $h_n\to h=\F^{-1}f$, and $\dot Q h_n\to \dot Q h$, the condition $h(a)=0$ can only be satisfied if the functions $h_n$ have support away from $a$.

Next we interpret the operators $-\Delta_\alpha$ as a perturbation of the maximal 
operator $-\Delta_0$, using the theory of singular perturbation developed in Section 1 of \cite{AK}, that we recall here briefly. 
Let $A$ be a self adjoint operator in the Hilbert space $\H$ as in Section \ref{s1.0}. Let $\gamma$ be a real number and $\a\in\H_{-2}$ with norm 1. A singular rank one perturbation of the operator $A$ is the operator defined by the following formula 
\[
A^\gamma=A+\gamma \a_c(\cdot) \a,
\]
where $\a_c$ is either $\a$ or a linear bounded extension of $\a$ \cite{AK} Section 1.3.2, depending whether $\a\in \H_{-1}-\H$ or $\a\in\H_{-2}-\H_{-1}$. A rigorous definition of this type of operators acting on the dual space of functionals $\H_{-2}$ has been given in \cite{AK} Section 1.3. The domain and the action of the operator are described in Theorems 1.3.1 and 1.3.2, respectively. 
Using this approach, we can define singular perturbed operators in the Hilbert space $\H$, by taking the restrictions of the operators $A^\gamma$ just defined (see equation (1.45) of \cite{AK} for the domain). We will use this definition and we will use the same notation. The operators defined in this way are self adjoint. 

If $\dot A$ is a symmetric operator in $\H$ with deficiency indices $(1,1)$, as in Section \ref{s1.0}, the self adjoint extension $A_\alpha$ of $\dot A$ described in Proposition \ref{l4}, with $\alpha={\rm ctg}\theta$ (or in Proposition \ref{l5}, when $\dot A=\dot Q$), coincides with the singular rank one perturbation 
$A^\gamma$ of the operator $A_0$ if 
\[
\frac{1}{\gamma}=\frac{1}{\alpha}-c,
\]
by Theorem 1.3.3 and the results of Section 1.3 of \cite{AK}, and where $c$ is a real number. 
Now, the mapping 
\[
\del:f\mapsto \int_{\Omega^n} \delta_{\Omega^n}(x-a) f(x) d^n x,
\]
defines a functional on $\H_2=W^{2,2}(\Omega^n)$, and it is easy to see 
that $\del\in \H_{-2}-\H_{-1}$ (see also \cite{AK} Section 1.5.1). 
However, $\|\del\|_{-2}\not=1$, thus we need to take in account a 
normalization factor and we define $\a=\frac{\del}{\|\del\|_{-2}}$
(note that $\|\del\|_{-2}\neq0$). Since
\[
\a(f)=\frac{f(a)}{\|\del\|_{-2}},
\]
the operator $-\Delta_\alpha$ corresponds to the  singular rank one perturbation of the operator $-\Delta_0=-\Delta^0$
\[
-\Delta^\gamma=-\Delta^0+\gamma \a_c(\cdot) \a
=-\Delta^0+\frac{\gamma}{\|\del\|_{-2}^2} \del_c(\cdot) \del,
\]
if we take 
$\frac{1}{\gamma}=\frac{1}{\alpha}-c,$
for any real $c$. If we compare this with the formal regularization  
$-\Delta^{g_R}=-\Delta+g_R\delta,$
of the formal perturbed operator $-\Delta^g=-\Delta^0+g\delta$ introduced 
in Section \ref{intro} (see equations (\ref{introeq1}) and 
(\ref{introeq3})), we get $g_R=\frac{\gamma}{\|\del\|_{-2}^2}$ and hence
\[
\frac{\|\del\|_{-2}^2}{\alpha}=\frac{1}{g_R}+c\|\del\|_{-2}^2.
\]

Summing up, we have proved that the regularized formal operator 
$-\Delta^{g_R}=-\Delta+g_R \delta$,
describing the Laplace operator with a delta type interaction considered 
in the introduction, corresponds to the operator 
$-\Delta^\gamma=-\Delta_{\alpha=\frac{\gamma}{1+c\gamma}}$, with $\gamma=\|\del\|^2_{-2}g_R$, and with any real $c$, and therefore it is unitary equivalent to the operator $Q_{\alpha=\frac{\gamma}{1+c\gamma}}$, defined explicitly in Sections \ref{ex1} and \ref{ex2},
respectively when $\Omega^n=\R^n$ or $\HH^n$. The resolvents are given by taking Fourier transform of the resolvents given in Lemmas \ref{ex1.l2} and  \ref{ex2.l2},
In all cases, the operator $-\Delta^0$ is the maximal operator, and the difference of the resolvents $R(\lambda,-\Delta^{g_R})-R(\lambda,-\Delta^0)$ is of trace class. 
The trace is given in Lemmas \ref{l9.1} and  \ref{l9.1b}.

In particular, we use this result in the formula for the  difference of 
the resolvents given in Proposition \ref{ex1.l20}. Since $\e$ is the 
Fourier transform of $\del$, 
$\|\del\|_{-2}^2=\|\e\|^2_{-2}=\|\psi_+\|^2_0=a_n$, we obtain 
\beq\label{pippo}
(R(\lambda,-\Delta^{g_R})-R(\lambda,-\Delta^0))\phi
   =\frac{1}{\frac{1}{g_R}+a_nc+b_n-g_n(\la)}
\int_{\Omega^n}\bar e(x)(R(\lambda,-\Delta^0)\phi)(x)d^nx \psi_\lambda.\\
\eeq

As observed in \cite{AK}, we have two free constants in this formula, and therefore a two parameters family of operators. While the constant $g_R$ has a physical meaning, since it is the coupling constant discussed in Section \ref{intro}, the constant $c$ should be fixed. However, a prescription to fix the constant $c$ has been introduced in Section 1.3.3 of \cite{AK} for the class of the {\em homogeneous operators}, defined as follows. Suppose there exists a group $G$ of unitary transformations of the Hilbert space $\H$. An operator $A$ is said to be {\em homogeneous} if it rescales in an appropriate way under the action of $G$, as in Lemma 1.3.2 of \cite{AK}. 
Now, suppose the operator $A^0$ is homogeneous accordingly to this definition. 
If this is the case, in the same lemma a condition is given for the existence 
of a singular rank one perturbation $A^\gamma$ of $A^0$, and its unicity is proved. The proof is based on the fact that the self adjoint operator $A^\gamma$ satisfies the same symmetry property as $A^0$ for one and only one value of the constant $c$. This  condition fixes the value of $c$. The case of the Laplace operator $-\Delta$ in $\R^3$ was discussed in Section 1.5.5 of \cite{AK}, where it is shown that $-\Delta$ is an homogeneous operator with respect to the group of the scaling transformations of $L^2(\R^3)$. We review this case and we also investigate the one dimensional case in the following Section \ref{sz2}. However, as observed at the end of Section 1.3.3 of \cite{AK}, in the 2-dimensional case, the Laplace operator is homogeneous but the condition for the existence of the singular rank one perturbation is not satisfied. It follows that singular rank one perturbations of the Laplace operator in dimension two do not exist. 

The situation is more difficult for the case of the Hilbert space $L^2(\HH^n)$, $n=1,2,3$. 
For in this case we do not have the group of symmetry given by scaling transformations, 
and consequently the prescription described above does not apply. 
On the other side, the situation is more delicate because of the following reason. If we compare the formula for the resolvent given in equation (\ref{pippo}) with the heuristic formula given in equation (\ref{introeq3}), a straightforward calculation shows that the two coincide if and only if we identify
\[
 \frac{1}{g_R} =\frac{1}{g_R}+a_n c+b_n.
\]

This condition can be satisfied either re-regularizing the coupling constant, or assuming the condition $c=-\frac{b_n}{a_n}$. Since the constants $a_n$ and $b_n$ depend on the geometry through the parameter $a$, the first possibility contradicts the plausible physical requirement that the coupling constant should not depend on the geometry. Therefore, we will assume the second possibility. With this choice, the self adjoint extension associated to the coupling constant $g_R$ is characterized in the following proposition and its corollary, whose proofs follow by the results of the previous sections.


\begin{prop}\label{lap1} The operator $-\Delta^{g_R}$ is the self adjoint extension 
of the non maximal multiplication operator $-\dot\Delta$ with  resolvent 
\begin{align*}
R(\lambda,-\Delta^{g_R})\phi&=R(\lambda,-\Delta^0)\phi
 +\frac{1}{\frac{1}{g_R}-g_n(\la)}
\int_{\Omega^n}\bar e(x)(R(\lambda,-\Delta^0)\phi)(x)d^nx \psi_\lambda,\\
\psi_\lambda(x)&=\frac{e(x)}{|x|^2-\lambda},
\end{align*}
where $\lambda\in\rho(-\Delta^{g_R})\cap\rho(-\Delta^0)$, and either 
$e(x)=\e^{iax}$ and the functions $g_n(\lambda)$ are given 
for $n=3,2,1$ and $\Omega^n=\R^n$ in
Section \ref{ex1}
or $e(x)=\sin(ax)$ and the $g_n(\lambda)$ are given for $n=3,2,1$ 
and $\Omega^n=\HH^n$ in Section \ref{ex2}.
We assume $g_Rg_n(\lambda)\neq1$, which corresponds to  pure continuum spectrum
(see Lemma \ref{l8b}).
\end{prop}

\begin{corol}\label{trace1} The difference of the resolvent $R(\lambda,-\Delta^{g_R})-R(\lambda,-\Delta^0)$ is trace class with trace
 \[
\Tr(R(\lambda,-\Delta^{g_R})-R(\lambda,-\Delta^0))=
-\frac{1}{\frac{1}{g_R}-g_n(\la)}\int_{\Omega^n}\frac{|e(x)|^2}{(|x|^2-\la)^2} d^nx.
\]

\end{corol}

\begin{rem}\label{rem08} In other words, we obtain formulas for the trace of the difference of the resolvents \\
$\Tr(R(\lambda,-\Delta^{g_R})-R(\lambda,-\Delta^0))$ simply by taking the correspondent formulas given in Lemmas \ref{l9.1} and \ref{l9.1b} and making the substitution 
$\frac{1}{g_R}=\frac{1}{\frac{a_n}{\alpha}+b_n}$, or 
$\frac{1}{g_R}=\frac{1}{\frac{\hat a_n}{\alpha}+\hat b_n}$, respectively.
\end{rem}

Explicit formulas for the trace for the cases of interest will be given in the following Section \ref{sz}.


We conclude this section by studying the eigenvalues of the operators $-\Delta^{g_R}$.

\begin{lem}\label{l8b} Let $-\Delta^{g_R}$ be the operator with resolvent given in Proposition  \ref{lap1}. Then, the point spectrum of $-\Delta^{g_R}$, 
$\Sp_p(-\Delta^{g_R})=\Sp_d(-\Delta^{g_R})$, is given as follows:
\begin{itemize}
\item if $\Omega^n=\R^3$ or $\Omega^n=\R^1$, then $\Sp_p(-\Delta^{g_R})=\emptyset$ if $g_R\geq 0$, while there is one negative eigenvalue otherwise;
\item if $\Omega^n=\HH^3$, then $\Sp_p(-\Delta^{g_R})=\emptyset$ if $a\geq -\frac{2g_R}{\pi^2}$, while there is one negative eigenvalue otherwise;

\item if $\Omega^n=\HH^1$, we distinguish two cases: if $g_R$ is finite, then $\Sp_p(-\Delta^{g_R})=\emptyset$ if $a\leq \frac{2g_R}{\pi}$, while there is one negative eigenvalue otherwise; if $g_R=\infty$, then $\Sp_p(-\Delta^{g_R})=\left\{\frac{\pi^2k^2}{a^2}\right\}_{k\in\Z_0}$.
\end{itemize}

\end{lem}

\noindent {Pooof.} Assume $\Im\sqrt{\lambda}\geq 0$. Consider first the case of $\Omega^n=\R^n$. Then, the first statement follow from Theorems 1.1.4 and 3.1.4 of \cite{AGHH}.

Next, consider the case of $\Omega^n=\HH^n$. If $n=3$, the possible eigenvalues are the solutions of the equation
\beq\label{es1}
\frac{1}{g_R}-\frac{\pi^2}{2} i \sqrt{\lambda}+\frac{\pi^2}{4a}\e^{2ia\sqrt{\lambda}}=0.
\eeq

Let $\sqrt{\lambda}=x+iy$. Then, equation (\ref{es1}) becomes
\[
b- i(x+iy)+\frac{\e^{2ia(x+iy)}}{2a}=b+y-ix+\frac{\e^{-2ay}}{2a}(\cos(2ax)+i\sin(2ax))=0,
\]
with $b=\frac{2}{\pi^2 g_R}$, and separating the real and imaginary parts
\beq\label{sis}
\left\{\begin{aligned}
&b+y+\frac{\e^{-2ay}}{2a}\cos(2ax)=0,\\
&x-\frac{\e^{-2ay}}{2a}\sin(2ax)=0.
\end{aligned}\right.
\eeq

Since $\lambda=x^2-y^2+2ixy$ must be real, $\sqrt{\lambda}=x+iy$ must be purely real or purely imaginary. Thus we look for solutions with $x=0$ or $y=0$.
In the first case, $x=0$, the system in equation (\ref{sis}) reduces to the equation
\[
y=-b-\frac{\e^{-2ay}}{2a},
\] 
that can be solved graphically with $f_1(y)=y$, $f_2(y)=-b-\frac{\e^{-2ay}}{2a}$. Since we have assumed $y\geq 0$, the existence of solutions depends on the value of $f_2(0)$. If $f_2(0)<0$, then there are no solutions. Since
\[
f_2(0)=-b-\frac{1}{2a},
\]
and this quantity is negative for all $a$ if $g_R\geq -\frac{2 a}{\pi^2}$, it follows that there are no solutions for all $a$ and $g_R\geq -\frac{2 a}{\pi^2}$, and there is one negative eigenvalue otherwise.. 
In the second case, $y=0$, and the system becomes
\[
\left\{\begin{aligned}
&b+\frac{1}{2a}\cos(2ax)=0,\\
&x-\frac{1}{2a}\sin(2ax)=0,
\end{aligned}\right.
\]
that has only the trivial solution, $x=0$ ($b=-2/2a$).

If $n=2$, the possible eigenvalues are the solutions of the equation
\beq\label{es2}
\frac{1}{g_R}-\frac{\pi}{4} \log(-\lambda)+\frac{\pi}{2}K_0(2a\sqrt{\lambda})=0.
\eeq

If $n=1$, the possible eigenvalues are the solutions of the equation
\beq\label{es3}
\frac{1}{g_R}-\frac{\pi i}{4 \sqrt{\lambda}}\left(1-\e^{2ia\sqrt{\lambda}}\right)=0.
\eeq

With $b=\frac{4}{g_R \pi}$, equation (\ref{es3}) becomes 
\[
ib\sqrt{\lambda}+1-\e^{2ia\sqrt{\lambda}}=0,
\]
that gives the system  
\beq\label{sis1}
\left\{\begin{aligned}
&1-by-\e^{-2ay}\cos(2ax)=0,\\
&bx-\e^{-2ay}\sin(2ax)=0.
\end{aligned}\right.
\eeq

With $x=0$, we obtain
\[
y=\frac{1}{b}-\frac{\e^{-2ay}}{b}.
\]

Since $f_2(y)=\frac{1}{b}-\frac{\e^{-2ay}}{b}$ has tangent with angular coefficient $f_2'(0)=\frac{2a}{b}$, the system in equation (\ref{sis1}) has one positive solution if and only if $a>\frac{b}{2}$.
With $y=0$, the system in equation (\ref{sis1}) becomes
\[
\left\{\begin{aligned}
&1-\cos(2ax)=0,\\
&bx-\sin(2ax)=0.
\end{aligned}\right.
\]

This system has only the trivial solution $x=0$ if $b\not=0$, and has infinitely many  solutions $x=\frac{\pi k}{a}$, $k\in\Z$, if $b=0$.

We point out that the spectrum in the case $\HH^1=[0,\infty)$, $g_R=\infty$, 
is as expected, since in this case the operator reduces to the sum 
of the Laplacian on the positive half line with Dirichlet boundary conditions, 
plus the Laplacian on the interval $[0,a]$.

\section{Determinant and partition functions}
\label{sz}

The aim of this section is to study the determinant of the operators described in 
Section \ref{lap}, and consequently to obtain explicit expression 
for the partition function of the associated models of the Casimir effect. 
We first recall how the zeta function regularization \cite{Haw} (see, also \cite{ZZ1} and references therein) is used to define the infinite determinants of  self adjoint positive operators $A$. In fact, one defines
\[
\det_\zeta A=\e^{-\zeta'(0,A)},
\]
where the zeta function of $A$ is by definition
\[
\zeta(s,A)=\sum_{\lambda\in \Sp A} \lambda^{-s},
\]
for $\Re(s)$ sufficiently large, and analytically continued elsewhere. 
Accordingly, we have for the the partition function $Z=(\det_\zeta \ell^2 A)^{-\frac{1}{2}}$,
\[
\log Z =\frac{1}{2}\zeta'(0,A)-\frac{1}{2}\zeta(0,A)\log \ell^2, 
\]
where $\ell$, a real non vanishing number, is the usual renormalization parameter.

More precisely, we need relative zeta functions and relative zeta determinants. We recall first the main definitions and some properties of relative zeta determinants in Section \ref{sz1}, and then we apply this method to the operators of interest in Section \ref{sz2}.

\subsection{Relative zeta determinant and relative partition function} \label{sz1} We will use the notation introduced in \cite{SZ} for relative zeta functions and we refer to that work or to the original paper of W. M\"{u}ller for more details \cite{Mul}.

Let $\H$ be a separable Hilbert space, and let $A$ and $A_0$ be two self adjoint non negative linear operators in $\H$. Suppose that $\Sp A=\Sp_c A$, namely that $A$ has a pure continuous spectrum. We recall that $R(\lambda,T)=(\lambda I-T)^{-1}$ denotes the resolvent of the operator $T$, and $\rho(T)$ the resolvent set.
Then, we introduce the following set of conditions on the pair $(A,A_0)$:
\begin{enumerate}
\item[(B.1)] the operator $R(\lambda,A)-R(\lambda,A_0)$ is of trace class for all $\lambda\in \rho(A)\cap\rho(A_0)$;
\item[(B.2)] as $\lambda\to \infty$ in $\rho(A)\cap\rho(A_0)$, there exists an asymptotic expansion of the form:
\[
\Tr (R(\lambda,A)-R(\lambda,A_0))\sim \sum_{j=0}^\infty \sum_{k=0}^{K_j} a_{j,k} (-\lambda)^{\alpha_j} \log ^k (-\lambda),
\]
where $-\infty<\dots<\alpha_1<\alpha_0$, $\alpha_j\to -\infty$, for large $j$, and $a_{j,k}=0$ for $k>0$;
\item[(B.3)] as $\lambda\to 0$, there exists an asymptotic expansion of the form
\[
\Tr (R(\lambda,A)-R(\lambda,A_0))\sim \sum_{j=0}^\infty b_{j} (-\lambda)^{\beta_j},
\]
where $-1\leq\beta_0<\beta_1<\dots$, and $\beta_j\to +\infty$, for large $j$.
\end{enumerate}

We introduce the further consistency condition 
\begin{enumerate}
\item[(C)] $\alpha_0<\beta_0$.
\end{enumerate}

It was proved in \cite{SZ} that if the pair of non negative self adjoint operators $(A,A_0)$ satisfies conditions (B.1)-(B.3), then it satisfies the conditions (1.1)-(1.3) of \cite{Mul}. In this situation we define the relative zeta function for the pair $(A,A_0)$ by the following equation
\beq\label{c2}
\zeta(s;A,A_0)=\frac{1}{\Gamma(s)}\int_0^\infty t^{s-1} \Tr\left(\e^{-tA_c}-\e^{-tA_0 }\right)dt,
\eeq
when $\alpha_0+1<\Re(s)<\beta_0+1$, and by analytic continuation elsewhere, and  we define the regularized relative determinant of the pair of operators $(A,A_0)$ by
\[
\det_\zeta (A,A_0)=\e^{-\frac{d}{ds}\zeta(s;A,A_0)\big|_{s=0}}.
\]

Introducing the relative spectral measure, we have the following useful representation  of the relative zeta function \cite{SZ}.

\begin{prop}\label{z1} Let $A$ be a non negative self adjoint operator and assume that there exists an operator $A_0$ such that the pair $(A,A_0)$ satisfies conditions (B.1)-(B.3), and (C). Then,
\[
\zeta(s;A,A_0) =\int_0^\infty v^{-2s}e(v;A,A_0) dv ,
\]
where the relative spectral measure is defined by
\begin{align*}
e(v;A,A_0)&=\frac{v}{\pi i}\lim_{\epsilon\to 0^+}\left(r(v^2\e^{2i\pi-i\epsilon};A,A_0)-r(v^2\e^{i\epsilon};A,A_0)\right),\\
r(\lambda;A,A_0)&=\Tr(R(\lambda,A)-R(\lambda,A_0)).
\end{align*}

The integral, the limit and the trace exist.
\end{prop}

Accordingly, we also define  the zeta
regularized partition function of a model described by the operator
$A$, under the assumption that there exists a second operator $A_0$
such that the pair of operators
$(A,A_0)$ satisfies assumptions (B.1)-(B.3), by
\beq\label{partition2} 
\log Z =\frac{1}{2}\zeta'(0;A,A_0)-\frac{1}{2}
\zeta(0;A,A_0)\log \ell^2, 
\eeq
where $\ell$, a real non vanishing number, is the usual renormalization parameter.

Next, we recall the main result of \cite{SZ} about the decomposition of the relative partition function of a finite temperature quantum field theory on an ultrastatic space time. Let $M$ be a smooth Riemannian manifold of dimension $n$, $n\in\N$,  and
consider the product $X=S^1_\frac{\beta}{2\pi}\times M$, where
$S^1_r$ is the circle of radius $r$. Let $\xi$ be a complex line
bundle over $X$, and $L$ a self adjoint non negative linear operator
in the Hilbert space $\H(M)$ of the $L^2$ sections of the
restriction of $\xi$ onto $M$, with respect to some fixed metric $g$
on $M$. Let $H$ be the self adjoint non negative operator
$H=-\p^2_u+L$, in the Hilbert space $\H(X)$ of the $L^2$ sections of
$\xi$, with respect to the product metric  $du^2\oplus g$  on $X$,
and with periodic boundary conditions on the circle. Assume that
there exists a second operator $L_0$ defined on $\H(M)$, such that
the pair $(L,L_0)$ satisfies the previous assumptions (B.1)-(B.3). Then, by Lemma
2.2 of \cite{SZ}, it follows that there exists a second operator $H_0$
defined in $\H(X)$, such that the pair $(H,H_0)$ satisfies those
assumptions too. Under these requirements, we introduce the relative
zeta regularized partition function of the model described by the
pair of operators $(H,H_0)$ using equation (\ref{partition2}), and
we have the following result \cite{SZ} Proposition 3.1.

\begin{prop}\label{pr1} Let $L$ be a non negative self adjoint operator on $M$, and $H=-\p^2_u+L$, on $S^1_r\times M$ as defined above. Assume there exists an operator $L_0$ such that the pair $(L,L_0)$ satisfies conditions (B.1)-(B.3). Then,
\begin{align*}
\zeta(0;H,H_0)&=
-\beta\Ru_{s=-\frac{1}{2}}\zeta(s;L,L_0),\\
\zeta'(0;H,H_0)&=-\beta\Rz_{s=-\frac{1}{2}}
\zeta(s;L;L_0)-2\beta(1-\log 2)\Ru_{s=-\frac{1}{2}}
\zeta(s;L,L_0)-2\log\eta(\beta ;L,L_0),
\end{align*}
where $H_0=-\p^2_u+L_0$, and the relative Dedekind eta function is
defined by
\begin{align*}
\log\eta(\tau;L,L_0)&=\int_0^\infty \log\big(1-\e^{-\tau v}\big) e(v;L,L_0) dv.
\end{align*}

The residues and the integral are finite.
\end{prop}

\subsection{Relative determinant for the Laplacian in the whole space} \label{sz2} The aim of this section, and of the following one, is to investigate the determinant and the partition function of the operator $-\Delta^{g_R}$ described in Section \ref{lap}, by means of the technique described in the previous section. More precisely, we consider in this section the operator acting in the space $L^2(\R^n)$, and in the following section the operator acting in the half-space  $L^2(\HH^n)$.  However, recalling  the analysis of Section \ref{lap}, singular rank one perturbations of the Laplace operator on $L^2(\R^n)$ according to the 
definition given in \cite{AK}, are well defined only for $n \neq 2$, $n\leq 3$. 
As a consequence, in the following, we shall restrict ourselves to the 
cases $n=1$ and $n=3$.

\begin{rem} The operator $-\Delta^{g_R}$ in the three dimensional space was originally described in \cite{BF} and more recently in \cite{AGHH} Section I.1, and in \cite{AK} Section 1.5. The operator $-\Delta^{g_R}$ in the one dimensional space was investigated in \cite{AGHH} Section I.3. In particular, it is worth to observe that the operators described by Albeverio \& others are obtained as rank one singular perturbations of the maximal operator $-\Delta^0$ (in the language of \cite{AK} and Section \ref{lap}) fixing the value of the free parameter $c$ by imposing the preservation of the symmetry under scaling transformations (as explained in Section \ref{lap}), while here we fix the value of the constant $c$ by the condition described at the end of Section \ref{lap}, namely $c=-\frac{b_n}{a_n}$. The two different prescriptions, however, define the same operator, as follows by comparing the formulas for the difference of the resolvents given in \cite{AGHH} Theorems I.1.1.2, and I.3.1.3, with the ones obtained here using Proposition \ref{lap1}, 
and Lemma \ref{ex1.l2}.
\end{rem}

The relative zeta function and the relative partition function for the operator $-\Delta^{g_R}$ in $\R^3$ have been evaluated in Section 4.1 of \cite{SZ}, and the following results were obtained:

\begin{align*}
\zeta(s;-\Delta^{g_R},-\Delta^0)
=&\frac{1}{2}\frac{(2\pi^2 g_R)^{2s}}{\cos\pi s},\\
\log\eta(\tau;-\Delta^{g_R},-\Delta^0)=&\log\Gamma\left(\frac{\tau}{4\pi^3 g_R}\right)+\frac{1}{2}\log \frac{\tau}{4\pi^3 g_R}-\frac{\tau}{4\pi^3 g_R}\left(\log \frac{\tau}{4\pi^3 g_R}-1\right)-\frac{1}{2}\log2\pi,\\
\log Z=&2\left(\log\frac{\ell}{2\pi^2 g_R}-1\right)\frac{\beta}{8\pi^3 g_R}-\log\eta\left(\beta;-\Delta^{g_R},-\Delta^0\right).\\
\end{align*}

For completeness, we investigate here the one dimensional case. 
The trace of the difference of the resolvents is given in Corollary 
\ref{trace1}, and using the results of Section \ref{ex1} (see also Remark \ref{rem08}) we obtain
\[
r(\lambda;-\Delta^{g_R},-\Delta^0)=\Tr(R(\lambda,-\Delta^{g_R})-R(\lambda,-\Delta^0))
=-\frac{1}{2\lambda(ib\sqrt{\lambda}+1)},
\]
where $b=\frac{1}{\pi g_R}$. The expansion for  large $\lambda$ is
\[
r(\la)=\frac{1}{2i b\,\la^{3/2}}+O\left(\frac{1}{\la^2}\right)\,,
\qquad\qquad\Im\sqrt\la>0\,,
\]
and for small $\lambda$
\[
r(\la)=-\frac{1}{2\la}+O(1)\,.
\]

It follows that all the conditions (B.1)-(B.3) of Section \ref{z1} are satisfied with $\alpha_0=-\frac{3}{2}$ or $\alpha_0=-1$, and $\beta_0=-1$. However, it should be noted that since $\alpha_0=-\frac{3}{2}$ if $b\not=0$, while $\alpha_0=-1$, if $b=0$, then 
condition (C) is satisfied when $b\not=0$, 
but is not satisfied when $b=0$. 
This is consistent with the fact that the limit case $b=0$ gives $g_R=\infty$, 
that corresponds to the limit case of the Laplacian 
with Dirichlet boundary condition at $x=0$, and a relative zeta function 
can not be defined in this case.

Next, we evaluate the relative spectral measure
\[
e(v;L,L_0)=\frac{v}{\pi i}\lim_{\epsilon\to 0^+}\left(r(v^2\e^{2i\pi-i\epsilon};L,L_0)-r(v^2\e^{i\epsilon};L,L_0)\right).
\]

We obtain (note that the spectral measure vanishes in both limit cases $b=0$ and $b=\infty$, corresponding to the Laplacian with Dirichlet boundary condition at $x=0$ and to the free Laplacian)
\[
e(v;-\Delta^{g_R},-\Delta^0)=-\frac{b}{\pi(1+b^2v^2)},
\]
and a simple calculation using the formula for the zeta function given in Proposition \ref{z1} gives
\[
\zeta(s;-\Delta^{g_R},-\Delta^0)=-\frac{(\pi g_R)^{-2s}}{2\sin(\pi s)}.
\]

Using the definition of the relative Dedekind eta function and equation (\ref{partition2}) for the partition function, we also obtain
\begin{align*}
\log\eta(\tau;-\Delta^{g_R},-\Delta^0)=&-\log\Gamma\left(\frac{g_R}{2}\tau\right)-\frac{1}{2}\log\frac{g_R}{2}\tau
+\frac{g_R}{2}\tau\left(\log \frac{g_R}{2}\tau-1\right)+\frac{1}{2}\log 2\pi,\\
\log Z=&-\frac{\pi}{4}g_R\beta-\log\eta(\beta;-\Delta^{g_R},-\Delta^0).\\
\end{align*}

\subsection{Relative determinant for the Laplacian in the half space} \label{sz3} We pass now to study the case of main interest, namely the operator $-\Delta^{g_R}$ acting in the space $L^2(\HH^n)$, $n=1$ and $3$. The case $n=2$ presents non trivial technical aspects that we are not able to tackle at the moment, and therefore its investigation is
postponed to a further occasion.

\subsubsection{The case $n=3$} The operator $-\Delta^{g_R}$ in $\HH^3$ is the operator with resolvent given in Proposition \ref{lap1} and corresponds to the Fourier transforms of the operator investigated in Section \ref{ex2}. In order to apply the results of Section \ref{sz1}, we first need to check the conditions (B.1)-(B.3). By Lemmas \ref{l7} and \ref{l8b},  the operator $-\Delta^{g_R}$ has pure continuous spectrum coinciding with the non negative real axis  for all $g_R\geq 0$. We will restrict ourselves to this case. Consider the pair $(-\Delta^{g_R}, -\Delta^0)$. By Corollary \ref{trace1} and Lemma \ref{l9.1b} (see also Remark \ref{rem08}), the difference of the resolvent is of trace class with trace
\begin{align}\nonumber
r(\lambda;-\Delta^{g_R},-\Delta^0)=\Tr(R(\lambda,-\Delta^{g_R})-R(\lambda,-\Delta^0))&=\frac{\pi^2 \left(1-\e^{2ai\sqrt{\lambda}}\right)}{2i\sqrt{\lambda}\left(\frac{1}{g_R}-\frac{\pi^2}{2} i \sqrt{\lambda}+\frac{\pi^2}{4a}\e^{2ia\sqrt{\lambda}}\right)}\\
\label{pip}&=\frac{1-\e^{2ai\sqrt{\lambda}}}{i\sqrt{\lambda}\left(b-i\sqrt{\lambda}+\frac{\e^{2ia\sqrt{\lambda}}}{2a}\right)},
\end{align}
where $b=\frac{2}{\pi^2 g_R}$, and $a$ is a real positive number that gives the position of the delta interaction. We obtain the expansions for large $|\lambda|$
\[
r(\la)=\frac{1}{\la}-\frac{ib}{\la^{3/2}}+O\left(\frac{1}{\la^2}\right)\,,
\qquad\qquad\Im\sqrt\la>0\,,
\]
and for small $|\lambda|$
\[
r(\la)=-\frac{4a^2}{1+2ab}-\frac{4a^3i}{1+2ab}\,\sqrt{\la}
-\frac{16a^4(1-ab)}{3(1+2ab)^2}\,\la+O(\la^{3/2})\,,
\]
and this shows that all the conditions (B.1)-(B.3) of Section  \ref{sz1}  are satisfied with $\alpha_0=-1<\beta_0=0$, and therefore also condition (C) is satisfied. 

Second, we evaluate the relative spectral measure
\beq\label{lpo}
e(v;L,L_0)=\frac{v}{\pi i}\lim_{\epsilon\to 0^+}\left(r(v^2\e^{2i\pi-i\epsilon};L,L_0)-r(v^2\e^{i\epsilon};L,L_0)\right).
\eeq

Substitution of the expression in equation (\ref{pip}) in equation (\ref{lpo}) gives
\[
e(v;-\Delta^{g_R},-\Delta^0)=-\frac{4a}{\pi}\frac{1-2ab+2ab\cos(2av)-2av\sin(2av)-\cos(2av)}{1+4a^2(b^2+v^2)+4ab\cos(2av)-4av\sin(2av)}.
\]

The function $e(v;-\Delta^{g_R},-\Delta^0)$ is a regular function of $v$ for all $v\geq 0$. For we show that there are no solution with $v\geq 0$ of the equation 
\[
1+4a^2(b^2+v^2)+4ab\cos(2av)-4av\sin(2av)=0.
\]

Consider the two curves:
\[
f_1(v)=4av\sin(2av)-4a^2 v^2,
\]
and
\[
f_2(v)=1+4a^2b^2+4ab\cos(2av),
\]
and assume $a,b,v\geq 0$. Obviously, $f_1(v)\leq f_3(v)$ where $f_3(v)=4av-4a^2 v^2$, is a parabola facing down with vertex $V=\left(\frac{1}{2a},1\right)$, that intercepts the horizontal axis in $v=0$ and $v=\frac{1}{a}$. 
On the other hand, 
\[
(1+2ab)^2\leq f_2(v)\leq (1-2ab)^2,
\]
and $f_2(v)$ oscillates around the value $1+4a^2b^2$, and $f_2(\frac{\pi}{4a})=1+4a^2b^2$.
This suggests to split the problem into the three intervals $\left[0,\frac{\pi}{4a}\right]$, $\left[\frac{\pi}{4a},\frac{1}{a}\right]$, and $\left[\frac{1}{a},\infty\right)$.
In the interval $\left[0,\frac{\pi}{4a}\right]$, $f_2$ is decreasing and therefore
\[
f_2(v)\geq f_2\left(\frac{\pi}{4a}\right)=1+4a^2b^2.
\]

On the other hand, in the same interval we have $f_3(v)\leq 1$
and therefore $f_3<f_2$  in this interval.
In the interval $\left[\frac{1}{a}, \infty\right)$, $f_3(v)\leq 0$, while $f_2(v)\geq (1-2ab)^2$, and the value is zero if and only if $v=\frac{\pi(1+2k)}{2a}$, $k\in\Z$. But we have 
$f_3(v)=0$ if and only if $v=0$ or $v=\frac{1}{a}$, and $\frac{1}{a}<\frac{\pi}{2a}$. Therefore, $f_3<f_2$ in this interval.
Eventually, consider the interval $\left[\frac{\pi}{4a}, \frac{1}{a}\right]$. In this interval, $f_2$ is decreasing and hence 
\[
f_2(v)\geq f_2\left(\frac{1}{a}\right)=1+4a^2b^2+4ab\cos 2.
\]

Also  $f_3$ is decreasing, and hence
\[
f_3(v)\leq f_2\left(\frac{\pi}{4a}\right)=\pi-\frac{\pi^2}{4}>0.
\]

Now, we can check that $1+4a^2b^2+4ab\cos 2>\pi-\frac{\pi^2}{4}$, 
for all $a$ and $b$, hence $f_3<f_2$ on this interval, and this concludes the proof that $e(v;-\Delta^{g_R},-\Delta^0)$ is a regular function of $v$ for all $v\geq 0$.

Third, we use Proposition \ref{z1} in order to obtain a suitable analytic extension of the relative zeta function. For we need the behavior for small and large $v$ of the function $e(v;-\Delta^{g_R},-\Delta^0)$. We have
\[
e(v;-\Delta^{g_R},-\Delta^0)=O(v^2),
\]
for $v\to 0^+$, and
\[
e(v;-\Delta^{g_R},-\Delta^0)=\frac{2\sin(2av)}{\pi v}
+\frac{2\sin^2(2av)}{a\pi v^2}-\frac{2(1-2ab)\sin^2(av)}{a\pi v^2}+O(v^{-3}),
\]
for $v\to +\infty$. So we decompose
\[
e(v;-\Delta^{g_R},-\Delta^0)=
e_0(v;-\Delta^{g_R},-\Delta^0)+e_\infty(v;-\Delta^{g_R},-\Delta^0),
\]
where
\begin{align*}
e_\infty(v;-\Delta^{g_R},-\Delta^0)&=\frac{2\sin(2av)}{\pi v}
      +\frac{2\sin^2(2av)}{a\pi v^2}-\frac{2(1-2ab)\sin^2(av)}{a\pi v^2},\\
e_0(v;-\Delta^{g_R},-\Delta^0)&=e(v;-\Delta^{g_R},-\Delta^0)
       -e_\infty(v;-\Delta^{g_R},-\Delta^0)\,,
\end{align*}
and 
\begin{align*}
\zeta(s;-\Delta^{g_R},-\Delta^0)&=\zeta_0(s;-\Delta^{g_R},-\Delta^0)+\zeta_\infty(s;-\Delta^{g_R},-\Delta^0)\\
&=\int_0^\infty v^{-2s}e_0(v;-\Delta^{g_R},-\Delta^0)dv+\int_0^\infty v^{-2s}e_\infty(v;-\Delta^{g_R},-\Delta^0)dv.
\end{align*}

Now $e_0(v;-\Delta^{g_R},-\Delta^0)$ goes to a constant for 
$v\to0$ and vanishes as $v^{-3}$ for $v\to\infty$, and so the function 
$\zeta_0(s;-\Delta^{g_R},-\Delta^0)$ 
is a regular function of $s$ in the interval 
$-1<\Re(s)<\frac{1}{2}$. 
The function $\zeta_\infty(s;-\Delta^{g_R},-\Delta^0)$ can be studied 
explicitly. We evaluate the integrals:
\[
\int_0^\infty v^{-2s-1} \sin (2av) dv=-(2a)^{2s}\sin(\pi s)\Gamma(-2s),
\]
for $-\frac12<\Re(s)<\frac12$ \cite{GZ} 3.761.4;
\[
\int_0^\infty v^{-2s-2} \sin^2 (av) dv=(2a)^{2s+1}\sin(\pi s)\Gamma(-2s-1),
\]
for $-\frac12<\Re(s)<\frac12$ \cite{GZ} 3.823. Collecting, we have
\[
\zeta_\infty(s;-\Delta^{g_R},-\Delta^0)=\frac{4}{\pi}(2a)^{2s}\sin(\pi s)\Gamma(-2s-1)(ab+s+2^{2s}).
\]

Thus we have the following representation for the relative 
zeta function when $-\frac12<\Re(s)<\frac{1}{2}$,
\[
\zeta(s;-\Delta^{g_R},-\Delta^0)=\frac{4}{\pi}(2a)^{2s}\sin(\pi s)\Gamma(-2s-1)(ab+s+2^{2s})+\int_0^\infty v^{-2s}e_0(v;-\Delta^{g_R},-\Delta^0)dv.
\]

This representation can be used in order to study the analytic continuation and in particular evaluate the residue and the finite part at $s=-\frac{1}{2}$. We obtain
\begin{align*}
\Ru_{s=-\frac{1}{2}}\zeta(s;-\Delta^{g_R},-\Delta^0)&=\frac{2}{\pi^2 g_R},\\
\Rz_{s=-\frac{1}{2}}\zeta(s;-\Delta^{g_R},-\Delta^0)&=\frac{1+\log 2}{\pi a}+\frac{2b(\gamma+\log(2a))}{\pi}+\zeta_0\left(-\frac{1}{2};-\Delta^{g_R},-\Delta^0\right)\\
&=\frac{1+\log 2}{\pi a}+\frac{4(\gamma+\log(2a))}{\pi^2 g_R}+\int_0^\infty v e_0(v;-\Delta^{g_R},-\Delta^0)dv.
\end{align*}

Using Proposition \ref{pr1} and the formula in equation (\ref{partition2}), we obtain the formula for the relative partition function 
\begin{align*}
\log Z=&-\frac\beta2\,\Rz_{s=-\frac{1}{2}}\zeta(s;-\Delta^{g_R},-\Delta^0)
      -\beta(1-\log(2\ell))\,\Ru_{s=-\frac{1}{2}}\zeta(s;-\Delta^{g_R},-\Delta^0)\\
       &-\log\eta(\beta;-\Delta^{g_R},-\Delta^0)\\
=&\frac{2\beta(\log (2\ell)-1)}{\pi^2 g_R}+\frac{\beta}{2}\left(\frac{1+\log 2}{\pi a}+\frac{4(\gamma+\log(2a))}{\pi^2 g_R}-\int_0^\infty v e_0(v;-\Delta^{g_R},-\Delta^0)dv\right)\\
&-\log\eta(\beta;-\Delta^{g_R},-\Delta^0).
\end{align*}

As a consequence, the vacuum energy of the system reads
\begin{align*}
E_c&=-\lim_{\beta\to\infty}\,\frac{\partial}{\partial_\beta}\,\log Z\\
    &=\frac{1+\log2}{2\pi a}+\frac{2}{\pi^2 g_R}\,\left(\gamma+1+\log\frac{a}{\ell}\right)
     +\frac12\,\int_0^\infty v e_0(v;-\Delta^{g_R},-\Delta^0)dv\,,
\end{align*}
since for large $\beta$ the exponential in the integral 
dominates in the definition of $\log\eta(\beta;-\Delta^{g_R},-\Delta^0)$.

We are interested in the behavior of the force 
$p=-\frac{\partial}{\partial a} E_c$ of the vacuum  
for small values of $a$. We need the expansion for small  $a$ of the integral
\begin{align*}
\zeta_0\left(-\frac{1}{2};-\Delta^{g_R},-\Delta^0\right)
   &=\int_0^\infty v^{-2s}e_0(v;-\Delta^{g_R},-\Delta^0)dv\\ 
   &=\frac{1}{a}\int_0^\infty \frac{x}{a} 
        e_0\left(\frac{x}{a};-\Delta^{g_R},-\Delta^0\right)dx,
\end{align*}
therefore we study the function
\[
f(x,a)=\frac{x}{a} e_0\left(\frac{x}{a};-\Delta^{g_R},-\Delta^0\right)
  =\frac{N(x,a)}{D(x,a)},
\]
where
\begin{align*}
N(x,a)=&-4x\sin x\left(\cos(5x)+4ab\cos(3x)+4a^2b^2\cos x\right)\\
&-2\sin x\left(2ab\sin(5x)+(1+8a^2b^2)\sin(3x)
                          +4ab(1-ab+2a^2b^2)\sin x\right),
\end{align*}
and
\[
D(x,a)=\pi x(1+4a^2b^2+4x^2+4ab\cos(2x)-4x\sin(2x))=\pi x g(x,a).
\]

This shows that the integral
\[
\int_0^\infty f(x,a) dx=\sum_{i} \int_0^\infty\frac{N_i(x,a)}{D(x,a)}dx,
\]
decomposes as a finite sum of terms, and in each term the numerator factors as
\[
N_i(x,a)=a^{p_i} h_i(x),
\]
where $p_i$ is $0,1,2$ or $3$, and the functions $h_i(x)$ are bounded. Thus, it remains to deal with the denominator. 
As a function of $a$, $g(x,a)$ is a parabola ``facing up'', 
so $g(x,a)\geq g(x,a_0)$, where $a_0$ is the vertex: 
so the solution of $\partial_a g(x,a)=8b^2 a+4b\cos(2x)=0$, i.e. 
$a_0=-\frac{\cos(2x)}{2b}$. Therefore
\[
D(x,a)\geq D(x,a_0)=\pi x\left(1+\frac{\cos^2(2x)}{2}+4x^2-4x\sin(2x)\right)>0,
\]
where it is easy to see that the function $D(x,a_0)$ is positive for all $x$. Thus, $|D(x,a)|>|D(x,a_0)|$, and
\[
\int_0^\infty |f(x,a)| dx\leq \sum_{i}\int_0^\infty\frac{a^{p_i}|h_i(x)|}{|D(x,a_0)|}.
\]

Now, it is also easy to see that
\[
\int_0^\infty \frac{a^{p_i}|h_i(x)|}{|D(x,a_0)|}<\infty,
\]
for all $i$, since $\frac{a^{p_i}|h_i(x)|}{|D(x,a_0)|}\sim \frac{1}{x^2}$ for each $i$.
This proves that the integral 
\[
\int_0^\infty \frac{x}{a} e_0\left(\frac{x}{a};-\Delta^{g_R},-\Delta^0\right)dx,
\]
converges uniformly for $a$ in compact sets, and therefore we can evaluate the behavior for small $a$ taking the expansion of the integrand for small values of $a$. We obtain 
\[
\zeta_0\left(-\frac{1}{2};-\Delta^{g_R},-\Delta^0\right)
    =\frac1a\,\left(I_0+abI_1+a^2b^2I_2\right)+O(a^2)\,.
\]

The integrals $I_n$ can be performed numerically. In particular we have
\begin{align*}
I_0&=-\frac{2}{\pi}\int_0^\infty\,\frac{\sin 
u(\sin(3u)+2u\cos(5u))}{u(1+4u^2-4u\sin(2u))}
   \,du\sim-0.12,\\
I_1&=-\frac{4}{\pi}\int_0^\infty\,
    \frac{\sin u\left(4u(1+4u^2)\cos(3u)-4u\cos u 
           -4u^2\sin(5u)+(1+16u^2)\sin u\right)}
         {u(1+4u^2-4u\sin(2u))^2}
\sim-0.51,\\
I_2&=-\frac{32}{\pi}\int_0^\infty\,\frac{u\sin u(8u^3\cos u-6u\cos(3u)-12u^2\sin u+\sin(5u))}
                         {(1+4u^2-4u\sin(2u))^3}
\,du\sim-1.04.
\end{align*}

This gives the behavior of the force for small $a$:
\[
p=-\frac{\partial}{\partial a}\,E_c
= \frac{1+\log2+2\pi I_0}{2\pi a^2}-\frac{2}{\pi^2g_R\,a}-\frac{2I_2}{\pi^4g_R^2}+
O(a).
\]

Using the numerical results given above, we see that for small values of $a$ 
the force is positive ($p\sim 0.15/a^2$).

\subsubsection{The case $n=1$} The operator $-\Delta^{g_R}$ in $\HH^1=[0,\infty)$ is the operator with resolvent given in Proposition \ref{lap1}, and corresponds to the Fourier transform of the operator investigated in Section \ref{ex2}. By Lemmas \ref{l7} and \ref{l8b},  the operator $-\Delta^{g_R}$ has pure continuous spectrum coinciding with the non negative real axis  if  $a\leq \frac{2g_R}{\pi}<\infty$. We will restrict our considerations to this case. 
Consider the pair $(-\Delta^{g_R}, -\Delta^0)$. By Corollary \ref{trace1} and Lemma \ref{l9.1b} (see also Remark \ref{rem08}), the difference of the resolvent is of trace class with trace
\begin{align*}
r(\lambda;-\Delta^{g_R},-\Delta^0)=\Tr(R(\lambda,-\Delta^{g_R})-R(\lambda,-\Delta^0))&=-\frac{\pi}{4i\lambda^\frac{3}{2}} \frac{1+\e^{2ai\sqrt{\lambda}}(2ia\sqrt{\lambda}-1)}{\frac{1}{g_R}+\frac{\pi}{4i\sqrt{\lambda}}(1-\e^{2ia\sqrt{\lambda}})}.\\
\end{align*}

We obtain the expansions for large values of $|\lambda|$
\[
r(\lambda;-\Delta^{g_R},-\Delta^0)=
\frac{ig_R\pi}{4\lambda^{3/2}}+O\left(\frac1{\lambda^2}\right)\,,
\qquad\qquad\Im\sqrt\lambda>0\,,
\]
and for small values of $|\lambda|$
\[
r(\lambda;-\Delta^{g_R},-\Delta^0)= 
\frac{i\,a^2g_R\pi}{(ag_R\pi-2)\,\sqrt\lambda}
  +\frac{a^3g_R\pi\,(ag_R\pi-8)}{3\,(ag_R\pi-2)^2}
     +O\left(\sqrt\lambda\right).
\]

This shows that all the conditions (B.1)-(B.3) of Section  
\ref{sz1}  are satisfied with $\alpha_0=-3/2<\beta_0=-1/2$, 
and therefore also condition (C) is satisfied. 
Second, we evaluate the relative spectral measure (see definition in Proposition \ref{z1}). We obtain
\[
e(v;-\Delta^{g_R},-\Delta^0)=\frac{4b\sin(av)}{\pi}\,\,
        \frac{(ab+1)\sin(av)-2av\cos(av)}
          {b^2+2v^2-b^2\cos(2av)-2bv\sin(2av)},
\]
where we have set $b=\frac{\pi g_R}{2}$. Proceeding as in the 
previous section, we 
show that the function $e(v;-\Delta^{g_R},-\Delta^0)$ is a regular function of $v$ for all $v\geq 0$. 
Third, we use Proposition \ref{z1} in order to obtain a suitable analytic extension of the relative zeta function. For we need the behavior for small and large values of $v$ of the function $e(v;-\Delta^{g_R},-\Delta^0)$. We have
\[
e(v;-\Delta^{g_R},-\Delta^0)=\frac{2a^2 b}{\pi(1-ab)}+O(v^2),
\]
for $v\to 0^+$, and
\[
e(v;-\Delta^{g_R},-\Delta^0)
  =\frac{2ab\sin(2av)}{\pi v}
   +\frac{2b\left(ab\sin^2(2av)-(1+ab)\sin^2(av)\right)}{\pi v^2}
    +O(v^{-3}),
\]
for $v\to +\infty$. So we decompose
\[
e(v;-\Delta^{g_R},-\Delta^0)=
      e_0(v;-\Delta^{g_R},-\Delta^0)+e_\infty(v;-\Delta^{g_R},-\Delta^0),
\]
where
\begin{align*}
e_\infty(v;-\Delta^{g_R},-\Delta^0)
    =&\frac{2ab\sin(2av)}{\pi v}
      +\frac{2b\left(ab\sin^2(2av)-(1+ab)\sin^2(av)\right)}{\pi v^2}\, ,\\
e_0(v;-\Delta^{g_R},-\Delta^0)=&
e(v;-\Delta^{g_R},-\Delta^0)-e_\infty(v;-\Delta^{g_R},-\Delta^0)\,,
\end{align*}
and 
\begin{align*}
\zeta(s;-\Delta^{g_R},-\Delta^0)&=\zeta_0(s;-\Delta^{g_R},-\Delta^0)+\zeta_\infty(s;-\Delta^{g_R},-\Delta^0)\\
&=\int_0^\infty v^{-2s}e_0(v;-\Delta^{g_R},-\Delta^0)dv+\int_0^\infty v^{-2s}e_\infty(v;-\Delta^{g_R},-\Delta^0)dv.
\end{align*}

As for the three-dimensional case, $e_0(v;-\Delta^{g_R},-\Delta^0)$
is a constant at $v=0$ and goes to zero as $v^{-3}$ at infinity. Then 
$\zeta_0(s;-\Delta^{g_R},-\Delta^0)$ is a regular function 
of $s$ in the interval $-1<\Re(s)<\frac12$. 

The function $\zeta_\infty(s;-\Delta^{g_R},-\Delta^0)$ can be studied 
explicitly.  The integrals involved are of the same type of the ones 
evaluated in the previous section. For $-\frac12<\Re (s)<\frac12$, we obtain
\[
\zeta_\infty(s;-\Delta^{g_R},-\Delta^0)
     =\frac{(2a)^{2s+1}b\left(2s+(2^{2s+1}-1)ab\right)}{\pi}\,\,
        \Gamma(-2s-1)\sin(\pi s).
\]

Thus we have the following representation for the relative zeta function 
when $-\frac12<\Re(s)<\frac12$,
\begin{align*}
\zeta(s;-\Delta^{g_R},-\Delta^0)
     =&\frac{(2a)^{2s+1}b\left(2s+(2^{2s+1}-1)ab\right)}{\pi}\,\,
        \Gamma(-2s-1)\sin(\pi s)\\
     &+\int_0^\infty v^{-2s}e_0(v;-\Delta^{g_R},-\Delta^0)dv.
\end{align*}

This representation can be used in order to study the analytic 
continuation and in particular evaluate the residue and 
the finite part at $s=-\frac{1}{2}$. We obtain
\begin{align*}
\Ru_{s=-\frac{1}{2}}\zeta(s;-\Delta^{g_R},-\Delta^0)
            &=-\frac{b}{2\pi}=-\frac{g_R}{4},\\
\Rz_{s=-\frac{1}{2}}\zeta(s;-\Delta^{g_R},-\Delta^0)
            &=\frac{b(1-\gamma-\log(2a)+ab\log2)}{\pi}
               +\zeta_0\left(-\frac{1}{2};-\Delta^{g_R},-\Delta^0\right)\\
            &=\frac{g_R(2-\gamma-2\log(2a)+ag_R\pi\log2)}{4}
               +\int_0^\infty v e_0(v;-\Delta^{g_R},-\Delta^0)dv.
\end{align*}

Using Proposition \ref{pr1} and the formula in equation (\ref{partition2}), we obtain the formula for the relative partition function 
\begin{align*}
\log Z=&-\frac\beta2\,\Rz_{s=-\frac{1}{2}}\zeta(s;-\Delta^{g_R},-\Delta^0)
      -\beta(1-\log(2\ell))\,\Ru_{s=-\frac{1}{2}}\zeta(s;-\Delta^{g_R},-\Delta^0)\\
       &-\log\eta(\beta;-\Delta^{g_R},-\Delta^0)\\
=& \frac{\beta g_R}{8}\,\left(2\gamma-ag_R\pi\log2+2\log\frac{a}{\ell}\right)
-\frac\beta2\,\int_0^\infty v e_0(v;-\Delta^{g_R},-\Delta^0)dv\\
&-\log\eta(\beta;-\Delta^{g_R},-\Delta^0).
\end{align*}

As a consequence, the vacuum energy of the system reads
\begin{align*}
E_c&=-\lim_{\beta\to\infty}\,\frac{\partial}{\partial_\beta}\,\log Z\\
  &= -\frac{g_R}{8}\,
       \left(2\gamma-ag_R\pi\log2+2\log\frac{a}{\ell}\right)
+\frac12\,\int_0^\infty v e_0(v;-\Delta^{g_R},-\Delta^0)dv\,,
\end{align*}
since for large $\beta$ the exponential function dominates 
in the function $\log\eta(\beta;-\Delta^{g_R},-\Delta^0)$.

We are interested in the behavior of the force 
$p=-\frac{\partial}{\partial a} E_c$ of the vacuum  
for small $a$. 
Therefore, we need an expansion for small values of $a$ of the integral
\[
\zeta_0\left(-\frac{1}{2};-\Delta^{g_R},-\Delta^0\right)
=\int_0^\infty ve_0(v;-\Delta^{g_R},-\Delta^0)dv.
\]

Proceeding as in the previous section, we show 
that the integral converges uniformly on compact subsets. 
Expanding for small values of $a$ we obtain
\[
\zeta_0\left(-\frac{1}{2};-\Delta^{g_R},-\Delta^0\right)
=-\frac{2\log 2}{\pi} ab^2+O(a^2)\,.
\]

This gives the behavior of the force for small values of $a$:
\[
p=-\frac{\partial}{\partial a}\,E_c
= \frac{g_R}{4a}+\frac{g_R^2\pi\log2}{8}+O(a).
\]

In this case the force is positive for small values of $a$.

\section{Appendix}
\label{sa}

\def\rl{ \sqrt\lambda \sgn(\Im\sqrt\lambda) }

\begin{lem}
\label{la1} Let $D^n_N$ be the closed disc of radius $N$ centered at the origin in $\R^n$, $n=1,2,3$. 
Then, for all $\lambda\in \C-(-\infty,0)$, and for large $N$ 
\begin{align*}
\int_{D^3_N} \frac{1}{|x|^2-\lambda}d^3x
     &=4\pi N+2\pi^2i \rl+o(1),\\
\int_{D^2_N} \frac{1}{|x|^2-\lambda}d^2x
     &=2\pi\log N-\pi\log(-\lambda)+o(1),\\
\int_{D^1_N} \frac{1}{|x|^2-\lambda}dx
     &=\frac{i\pi}{\rl}-\frac{2}{N}+o\left(\frac{1}{N}\right).
\end{align*}
\end{lem}

\noindent {Pooof.} 
Using polar coordinates we have
\begin{align*}
\int_{D^3_N}\frac{1}{|x|^2-\lambda}d^3x
     &=4\pi \int_0^N\frac{r^2}{r^2-\lambda} dr\\
     &=4\pi N+4\pi\lambda \int_0^N\frac{dr}{r^2-\lambda}
=4\pi N+2\pi^2i\rl+o(1).
\end{align*}

In a similar way we get corresponding results for $n=2$ and $1$.

\begin{lem}\label{la2}
For $n=1,2,3$ we have
\[
\int_{\R^n}\frac{d^nx}{\left||x|^2\pm i\right|^2}=
\left\{\begin{array}{ll} 
    \sqrt{2}\pi^2, & n=3,\\   
    \frac{\pi^2}2, & n=2,\\   
    \frac{\pi^2}{\sqrt2}, & n=1.
\end{array}
\right.
\]
\end{lem}
\noindent {Pooof.}
We observe that 
\[
\frac{1}{\left||x|^2\pm i\right|^2}=\frac{1}{|x|^4+1}=
\frac{1}{2i}\left(\frac{1}{|x|^2+i}-\frac{1}{|x|^2-i}\right).
\]

Then the results follow as a consequence of Lemma \ref{la1}.

\begin{lem}
\label{la1b} Let $B^n_N$ be the closed half disc of radius $N$ centered at the origin in $\R^n$, $n=1,2,3$. 
Then, for all $\lambda\in \C-(-\infty,0)$, $a\in\R^n$ and large $N$
\begin{align*}
\int_{B^3_N} \frac{\sin^2(ax)}{|x|^2-\lambda}d^3x
    &=\pi N+\frac{i\pi^2}{2}\rl-\frac{\pi^2}{4a} \e^{2ia\rl}+o(1),\\
\int_{B^2_N} \frac{\sin^2(ax)}{|x|^2-\lambda}d^2 x
    &=\frac\pi2 \log N-\frac\pi4 \log(-\lambda)+\frac\pi2 K_0(2ai\sqrt{\lambda})+o(1),\\
\int_{B^1_N} \frac{\sin^2(ax)}{|x|^2-\lambda}d x
    &=\frac{i\pi}{4\rl}\left(1-\e^{2ia\rl}\right)-\frac{1}{2N}+o\left(\frac{1}{N}\right),
\end{align*}
where $K_0(z)$ is a Bessel function.
\end{lem}

\noindent {Pooof.} (See also \cite{GZ} 3.723.3 and 3.723.10). 
First of all we observe that
\begin{align*}
\int_{B^n_N}\frac{\sin^2(ax)}{|x|^2-\lambda}d^n x
=\frac{1}{2}\int_{D^n_N}\frac{\sin^2(ax)}{|x|^2-\lambda}d^n x
=\frac{1}{4}\int_{D^n_N}\frac{1}{|x|^2-\lambda}d^n x
   -\frac{1}{4}\int_{D^n_N}\frac{\cos(2ax)}{|x|^2-\lambda}d^n x.
\end{align*}

Now we see that for $n=1,2,3$ the first integral in the latter line  
has been already computed in Lemma \ref{la1}, while for the second one,
by taking polar coordinates and putting $a$ on the positive $z$-axis we get ($n=3$)
\begin{align*}
\int_{D^3_N}\frac{\cos(2ax)}{|x|^2-\lambda}d^2x
 &=2\pi\int_{r=0}^N\int_{u=-1}^1\frac{r^2\cos(2aru)}{r^2-\lambda}du dr\\
 &=\frac{2\pi}a\int_{r=0}^N\frac{r \sin(2ar)}{r^2-\lambda} dr
    =\frac{\pi^2}{a} e^{2ia\rl}+o(1),
\end{align*}
while for $n=2,1$ we obtain respectively
\begin{align*}
\int_{D^2_N}\frac{\cos(2ax)}{|x|^2-\lambda}d^2x
=-2\pi K_0(2ai\sqrt{\lambda})+o(1),
\end{align*}
\begin{align*}
\int_{-N}^N\frac{\cos(2ax)}{x^2-\lambda}dx
=\frac{\pi i e^{2ia\rl}}{\rl}
+o(1/N),
\end{align*}
$K_0(z)$ being a Bessel function. The thesis follows from these results and Lemma \ref{la1}.

By an easy computation we get:

\begin{lem}\label{aa1}
For $n=1,2,3$, $a\in\R^n$ and $\Im(\sqrt\lambda)>0$ we have
\[
\int_{\R^n}\frac{d^nx}{(|x|^2-\lambda)^2}=
\left\{\begin{array}{ll} 
    \frac{i\pi^2}{\sqrt\lambda}, & n=3,\\   
    -\frac{\pi}{\lambda}, & n=2,\\   
    -\frac{i\pi}{2\lambda^{3/2}}, & n=1.
\end{array}
\right.
\]
\end{lem}

\begin{lem}\label{aa2}
For $n=1,2,3$, $a\in\R^n$ and $\Im(\sqrt\lambda)>0$ we have
\[
\int_{\R^n}\frac{\sin^2(ax)}{(|x|^2-\lambda)^2} d^nx=
\left\{\begin{array}{ll} 
    \frac{i\pi^2}{2\sqrt\lambda}\left(1-e^{2ia\sqrt\lambda}\right),& n=3,\\   
    -\frac{\pi}{2\lambda}-\frac{\pi i a}{\sqrt\lambda} K_1(-2ia\sqrt\lambda), & n=2,\\   
    -\frac{i\pi}{4\lambda^{3/2}}\left(1-\e^{2ia\sqrt\lambda}\right)
      +\frac{\pi a}{2\lambda} \e^{2ia\sqrt\lambda}, & n=1.
\end{array}
\right.
\]
where $K_1(z)$ is a Bessel function.
\end{lem}

\noindent {Pooof.}
As in Lemma \ref{la1b} we write
\[
\int_{\R^n}\frac{\sin^2(ax)}{(|x|^2-\lambda)^2} d^nx
   =\frac12 \int_{\R^n}\frac{1}{(|x|^2-\lambda)^2} d^nx-
     \frac12 \int_{\R^n}\frac{\cos(2ax)}{(|x|^2-\lambda)^2} d^nx.
\]

Then, choosing polar coordinates and putting $a$ along the positive $z$-axis, 
for $n=3$ we get
\begin{align*}
\int_{\R^3}\frac{\cos(2ax)}{(|x|^2-\lambda)^2} d^3x
  &=2\pi\int_{r=0}^\infty\int_{u=-1}^1\frac{r^2\cos(2aru)}{(r^2-\lambda)^2}du dr\\
    &=\frac{2\pi}a\int_{r=0}^\infty\frac{r \sin(2ar)}{(r^2-\lambda)^2} dr
      =\frac{i\pi^2}{\sqrt\lambda} e^{2ia\sqrt\lambda}.
\end{align*}

In a similar way we obtain
\begin{align*}
\int_{\R^2}\frac{\cos(2ax)}{(|x|^2-\lambda)^2} d^2x
   &=\frac{2\pi i a}{\sqrt\lambda} K_1(-2ia\sqrt\lambda),\\
\int_{\R}\frac{\cos(2ax)}{(x^2-\lambda)^2} dx
   &=-\left(\frac{\pi a}{\lambda}+\frac{i\pi}{2\lambda^{3/2}}\right) \e^{2ia\sqrt\lambda}.
\end{align*}

The thesis follows from these results and Lemma \ref{aa1}.

\end{document}